\documentclass[sigconf]{acmart}

\AtBeginDocument{%
  }

\copyrightyear{2026}
\acmYear{2026}
\setcopyright{cc}
\setcctype{by}
\acmConference[MM '26]{Proceedings of the 34th ACM International Conference on Multimedia}{November 10--14, 2026}{Rio de Janeiro, Brazil}
\acmBooktitle{Proceedings of the 34th ACM International Conference on Multimedia (MM '26), November 10--14, 2026, Rio de Janeiro, Brazil}
\acmDOI{10.1145/3767308.3836500}
\acmISBN{979-8-4007-2213-4/2026/11}
\usepackage{algorithm}
\usepackage{algorithmic}
\usepackage{subcaption}
\usepackage{multirow}

\begin{document}
\newcommand{\mascot}{\mbox{\textsc{MASCOT}}}
\newcommand{\extver}{the appendix}

\title{MASCOT: Model-Aware Submodular Coverage for Composite-Attribute Text-to-Image Retrieval}
\author{Aaryan Sharma}
\affiliation{%
  \institution{Indian Institute of Technology Bombay}
  \department{Department of Electrical Engineering}
  \city{Mumbai}
  \country{India}
}

\author{Vishak Prasad C}
\affiliation{%
  \institution{Indian Institute of Technology Bombay}
  \department{Department of Computer Science and Engineering}
  \city{Mumbai}
  \country{India}
}

\author{Virendra Singh}
\affiliation{%
  \institution{Indian Institute of Technology Bombay}
  \department{Department of Electrical Engineering}
  \city{Mumbai}
  \country{India}
}

\author{Ganesh Ramakrishnan}
\affiliation{%
  \institution{Indian Institute of Technology Bombay}
  \department{Department of Computer Science and Engineering}
  \city{Mumbai}
  \country{India}
}

\renewcommand{\shortauthors}{Aaryan Sharma, Vishak Prasad C, Virendra Singh, and Ganesh Ramakrishnan}
\begin{abstract}

Vision-Language Models (VLMs) are highly effective in retrieving semantically relevant images. However, in practice, relevance alone is often insufficient. Systems must also achieve Result Diversification (RD) across composite attributes such as geography and time, a task for which precise control remains challenging. Current re-ranking methods, such as Multi-Source Determinantal Point Processes (MS-DPP), address this using manifold-based repulsion over similarity representations. Although this strategy is effective for broad exploration, it exposes a key limitation in manifold-based models: when subjected to diversity-decrease tasks on discrete metadata, they suffer substantial degradation in early-rank recall.

To bridge this gap, we introduce \mascot{} (Model-Aware Submodular Coverage for Composite-Attribute Text-to-Image Retrieval). Instead of relying on manifold repulsion, \mascot{} formulates multi-attribute diversity as a resource allocation problem, projecting attributes into a soft-binning space weighted by query-driven importance. Averaged across the three PixelProse diversity-decrease tasks, \mascot{} preserves an early-rank recall (R@10) of 88.58\%, while MS-DPP retains 67.63\%. The margin widens under composite constraints: on \texttt{PP\_geo\_hour}, where temporal and geographic diversity must be suppressed simultaneously, MS-DPP's recall collapses from 0.9737 to 0.4931 and its top-ranked result degrades to R@1 = 0.23, while \mascot{} holds R@10 = 0.9410 and R@1 = 0.7202 at a diversity metric above the unconstrained baseline. We do not claim uniform superiority: on aggregate diversity--relevance scores our own simpler ablations attain higher harmonic means on all three decrease tasks, and \mascot{}'s advantage is specific to recall beyond rank~1 under composite constraints.\footnote{We release our code and supplementary materials at \url{https://github.com/AaryanSharma/MASCOT}.}

\end{abstract}




\maketitle

\section{Introduction}

Vision-Language Models (VLMs) \cite{li2023blip2bootstrappinglanguageimagepretraining,radford2021learningtransferablevisualmodels,jia2021scalingvisualvisionlanguagerepresentation,NEURIPS2021_50525975,alayrac2022flamingovisuallanguagemodel,yu2022cocacontrastivecaptionersimagetext} have established a new paradigm in Text-to-Image Retrieval, demonstrating remarkable accuracy in zero-shot semantic matching. However, in practical deployment, retrieving the most semantically relevant images is often insufficient. Modern search systems require Result Diversification (RD) \cite{Kulesza_2012,10.1145/290941.291025,10.1007/978-3-642-12275-0_11,10.1145/1498759.1498766,capannini2011efficientdiversificationwebsearch}, the ability to control the distribution of retrieved images across composite attributes such as geographic location and temporal metadata. This challenge has recently been formalized as the Contextual Diversity Refinement of Composite Attributes (CDR-CA) \cite{sogi2025msdppsmultisourcedeterminantalpoint}.

The prevailing paradigm for post-hoc Result Diversification relies on Determinantal Point Processes (DPPs) \cite{Kulesza_2012,lyons2003determinantalprobabilitymeasures,Lavancier_2014}, which model item selection through determinantal repulsion. Recent state-of-the-art approaches, such as Multi-Source DPP (MS-DPP) \cite{sogi2025msdppsmultisourcedeterminantalpoint}, extend this framework to handle multiple metadata attributes by interpolating similarity matrices on the tangent space of the Symmetric Positive Definite (SPD) manifold. Extensions of DPPs have been explored for scalable inference, mini-batch diversification, and recommendation systems \cite{chen2018fast,zhang2017determinantalpointprocessesminibatch,mariet2019dppnetapproximatingdeterminantalpoint}. Although this manifold-based repulsion is highly effective for broad exploration (pushing visually and temporally similar images apart to maximize diversity), it introduces a critical, previously unaddressed vulnerability. 

In this work, we expose a failure mode in modern DPP-based retrievers \cite{sogi2025msdppsmultisourcedeterminantalpoint}: they are ill-equipped for \emph{attribute-specific redundancy enforcement}, the task of tightly concentrating retrieved results within a narrow target range of one or more discrete metadata attributes (e.g., images from a specific geographic region or a narrow temporal window) while preserving semantic relevance to the query. Unlike diversity increase, which aligns naturally with the repulsive geometry of DPPs, diversity decrease demands that the model suppress variation along a target attribute axis, a constraint that continuous manifold repulsion cannot satisfy without discarding semantically relevant images. Averaged over the three PixelProse diversity-decrease tasks, MS-DPP retains an early-rank recall (R@10) of 0.6763, against 0.8858 for the coverage-based formulation we propose. The gap widens sharply under composite constraints: on \texttt{PP\_geo\_hour}, where temporal and spatial diversity must be suppressed simultaneously, MS-DPP's recall falls from 0.9737 to 0.4931 and its overall score reaches only 0.3850. Tangent normalization mitigates the recall loss but does not resolve the underlying trade-off: the fully normalized MS-DPP+TN+TVMS variant recovers recall to 0.9021 while attaining an even lower overall score (0.3363), compressing diversity only marginally beyond the unconstrained baseline.

To address this limitation, we propose \mascot{}. Instead of relying on spatial repulsion, \mascot{} formulates multi-attribute diversification as a dynamic resource allocation problem over discrete conceptual buckets, termed \textbf{Information Units (IUs)}, each representing a target boundary such as a one-hour window or a geographic grid cell. To avoid artificial boundary penalties, treating an image taken at 12:59~PM as disjoint from one at 1:01~PM, \mascot{} projects boundary-sensitive metadata into a probabilistic soft-binning space and weights each IU by query-driven semantic importance so that the retrieval budget is spent only on bins containing relevant content.

The primary contributions of this paper are threefold:
\begin{itemize}
    \item \textbf{Identification of Manifold Vulnerability:} We empirically demonstrate that state-of-the-art manifold repulsion techniques (MS-DPP) \cite{sogi2025msdppsmultisourcedeterminantalpoint} degrade sharply during diversity-decrease tasks on discrete metadata, sacrificing substantial early-rank recall on composite-attribute constraints.
    \item \textbf{The \mascot{} Framework:} We introduce a submodular coverage framework that models Contextual Diversity Refinement as a probabilistic resource allocation problem. Because its query-driven bin weights are derived from relevance, the first selection is governed by two correlated terms; we characterize when this preserves the top-ranked result and when concentration displaces it (\extver{}).
    \item \textbf{Semantic-Safe Diversity Decrease:} Across all three PixelProse diversity-decrease tasks, \mascot{} retains higher early-rank recall than MS-DPP (mean R@10 0.8858 vs.\ 0.6763), with the margin widening from 0.04 on single-attribute geography to 0.45 on the composite \texttt{PP\_geo\_hour} task, where it holds R@1 = 0.7202 against MS-DPP's 0.2346. We further show that this advantage is confined to ranks $K \geq 3$ under composite constraints: at rank~1, and on single-attribute tasks throughout, our simpler Uniform Binning ablation is competitive or better (Section~\ref{sec:results}).
    
\end{itemize}

\section{Related Work}
\label{sec:related}
\subsection{Text-to-Image Retrieval}
Text-to-image retrieval aligns visual and textual modalities in a shared embedding space. Learning objectives broadly fall into dual-encoder approaches \cite{radford2021learningtransferablevisualmodels,jia2021scalingvisualvisionlanguagerepresentation,yuan2021florencenewfoundationmodel,faghri2018vseimprovingvisualsemanticembeddings}, which embed images and text independently and align them with contrastive or ranking losses, and fusion-based approaches \cite{NEURIPS2021_50525975,chen2020uniteruniversalimagetextrepresentation}, which predict a matching score from jointly encoded pairs. Cross-attention architectures \cite{lee2018stackedcrossattentionimagetext} construct finer-grained region--word alignments at higher computational cost, motivating lighter convolution-free designs \cite{kim2021viltvisionandlanguagetransformerconvolution}. While these dense retrievers achieve strong semantic accuracy, they tend to return visually homogeneous results, motivating post-hoc diversification. A fuller treatment is given in \extver{}.

\subsection{Result Diversification}

Result Diversification (RD) counteracts search redundancy by balancing semantic relevance with item novelty. Early heuristic approaches such as Maximal Marginal Relevance (MMR) \cite{10.1145/290941.291025}  use greedy sequential selection to penalize similarity to previously selected items. Subsequent work introduced intent-aware diversification frameworks such as xQuAD \cite{10.1145/1772690.1772780,10.1007/978-3-642-12275-0_11}, PM-2 \cite{10.1145/2348283.2348296}, and IASelect \cite{10.1145/1498759.1498766}, which explicitly model query ambiguity and subtopics. Comprehensive field surveys highlight the evolution of these metrics from simple heuristics to complex probabilistic objectives \cite{10.1561/1500000040,wu2024resultdiversificationsearchrecommendation}. Efficient diversification methods for large-scale retrieval have also been explored \cite{capannini2011efficientdiversificationwebsearch}. To provide more rigorous probabilistic modeling of diversity, Determinantal Point Processes (DPPs) were adapted for machine learning \cite{Kulesza_2012,10.5555/3104482.3104632}. DPPs leverage spatial repulsion to model negative correlations, though they often face computational scalability bottlenecks due to exact inference requirements \cite{chen2018fast}. Recent methodologies evaluate the success of these diversification algorithms using advanced and reference-free metrics such as the Vendi Score \cite{friedman2023vendiscorediversityevaluation}. The Vendi Score measures the effective number of unique items by calculating the exponential of the von Neumann entropy of a similarity matrix, making it highly effective for identifying mode collapse and redundancy.

\subsection{Contextual Diversity Refinement and Submodularity}

Modern search systems increasingly require precise control over specific discrete metadata attributes. This challenge is formalized as the Contextual Diversity Refinement of Composite Attributes (CDR-CA) \cite{sogi2025msdppsmultisourcedeterminantalpoint}, which aims to refine the diversities of multiple attributes simultaneously according to application-specific contexts. Recent approaches like Multi-Source DPPs (MS-DPP) \cite{sogi2025msdppsmultisourcedeterminantalpoint} address this by interpolating distinct attribute similarity matrices on the tangent space of the Symmetric Positive Definite manifold. These methods are grounded in Determinantal Point Processes (DPPs), which model diversity through global negative correlations and repulsive interactions between items \cite{Kulesza_2012}. Extensions of DPPs to continuous and manifold-structured spaces further reinforce this repulsive modeling paradigm \cite{affandi2013approximateinferencecontinuousdeterminantal}. Beyond retrieval, DPP-based repulsion has also been applied to promote diversity in other machine learning settings, such as mini-batch selection for stochastic optimization \cite{zhang2017determinantalpointprocessesminibatch}. Recent work has also explored learning-based approximations of DPPs, such as DPPNet, which models repulsive subset selection using deep neural architectures \cite{mariet2019dppnetapproximatingdeterminantalpoint}.

Submodular optimization provides a principled framework for modeling diminishing returns in subset selection problems \cite{nemhauser1978analysis,dughmi2011submodularfunctionsextensionsdistributions,bian2017guaranteed}. Submodular functions admit efficient greedy maximization with $(1 - 1/e)$ approximation guarantees. These properties have been widely exploited in document summarization \cite{lin2011class}, sensor placement \cite{1662434}, and data subset selection \cite{10.1145/1281192.1281239,pmlr-v37-wei15}. Furthermore, researchers have expanded submodular maximization to handle complex constraints and massive datasets, introducing techniques like stochastic greedy approximations \cite{mirzasoleiman2015lazier}, knapsack constraints \cite{iyer2013submodularoptimizationsubmodularcover}, and bridging submodularity with spectral analysis for regression \cite{das2011submodularmeetsspectralgreedy}.

In subset-selection problems more broadly, submodularity has been used to model coverage and diversity simultaneously \cite{yehuda2022active,chen2024less}. Importantly, DPP objectives can be interpreted as log-submodular functions, establishing a deep connection between probabilistic repulsion and combinatorial optimization \cite{Kulesza_2012}.

Coverage-based selection over multi-attribute data was formalized by Xu et al.\ \cite{xu2014efficient}, whose Information Unit vocabulary we adopt. Their framework attaches a weight $w_u$ to each unit, but this weight is a static property of the unit, fixed prior to selection; our $\Omega(u,q)$ is computed per query from the peak relevance the unit contains, so the same Information Unit carries different value under different queries. Alignment-based relevance scoring has separately been used for CLIP-guided data selection \cite{yang2025a}, which combines an image--text alignment score with a local-density diversity score under a budget constraint; that formulation scores samples independently, so diversity enters only through a density prior, whereas coverage makes the marginal value of a sample depend on what has already been selected.

Bidirectional control itself is not new. MS-DPP \cite{sogi2025msdppsmultisourcedeterminantalpoint} attaches a direction indicator $s_j \in \{-1,+1\}$ to each attribute's tangent vector on the SPD manifold before unification, so that a negative sign reverses the determinantal repulsion for that attribute. \mascot{} instead realizes bidirectionality within a coverage formulation, where $d \in \{-1,+1\}$ converts marginal coverage into a penalty. The distinction is the saturation behavior analyzed in Section~\ref{sec:theory}: the coverage penalty is bounded by the residual capacity of a bin, so the penalty contributed by a bin decays to zero as that bin saturates, whereas determinantal repulsion has no such floor and continues to penalize similar selections as the set grows.

\section{Vulnerability of Manifold Repulsion}

The current state-of-the-art framework for this task is the Multi-Source Determinantal Point Process (MS-DPP) \cite{sogi2025msdppsmultisourcedeterminantalpoint}. MS-DPPs model the selection process by defining a unified similarity matrix $M$ on the Symmetric Positive Definite (SPD) manifold:
\begin{equation}
    M = \text{expm}\left( \sum_{j=1}^{N_A} s_j w_j \text{logm} S_{j} \right)
\end{equation}
where $S_j$ is the similarity matrix for the $j$-th attribute, $w_j$ is the user-defined weight, and $s_j \in \{-1, +1\}$ dictates whether to increase ($+1$) or decrease ($-1$) diversity. To prevent attributes with large tangent vector norms from dominating the selection, MS-DPP introduces Tangent Normalization (TN), which rescales each attribute's tangent vector to the norm of the relevance matrix's tangent vector; a second stage applies the same rescaling to the unified tangent vector.

While mathematically elegant for increasing diversity ($s_j = +1$), this formulation introduces a vulnerability during diversity-decrease tasks. When $s_j = -1$, the $j$-th attribute's tangent vector is negated before unification, so the unified kernel rewards selections that are mutually similar along that attribute. Because determinantal selection scores an item by its marginal contribution to the volume of the selected parallelepiped, the penalty for distinctness admits no saturation floor. An image remains penalized for being semantically distinct even after the target attribute range is already densely covered. Under composite constraints, where this pressure applies along multiple attribute axes simultaneously, the model discards semantically relevant images to satisfy the geometric objective, producing the early-rank recall degradation observed in Section~\ref{sec:results}.

\section{The \mascot{} Framework}
\label{sec:framework}

We formalize the Contextual Diversity Refinement of Composite Attributes (CDR-CA) task as a subset selection problem. Given a text query $q$, a base Vision-Language Model (VLM) retrieves a candidate set $\mathcal{V}$ of $N$ images. Each image $i \in \mathcal{V}$ has a raw semantic relevance score $r_i \in [0, 1]$ and a set of discrete metadata attributes $\mathcal{A} = \{a_{\text{time}}, a_{\text{geo}}\}$. The objective is to select a subset $\mathcal{S} \subset \mathcal{V}$ of size $K$ that optimizes a trade-off between semantic relevance and a context-specific multi-attribute diversity distribution.

To overcome the limitations of manifold repulsion, we propose \mascot{}. Instead of modeling diversity as continuous spatial repulsion, \mascot{} frames it as a probabilistic resource allocation problem over discrete metadata boundaries. 

\subsection{Soft Information Units (IUs)}

We define a discrete set of Information Units, $\mathcal{U}$ \cite{xu2014efficient}. For temporal tasks, $|\mathcal{U}| = 24$ (hourly bins); for geographic tasks, $\mathcal{U}$ is a spatial grid. Because discrete metadata are boundary-sensitive (e.g., an image taken at 12:59 PM is semantically identical to one at 1:01 PM, despite crossing an hourly boundary), we transform the deterministic metadata $a_i$ into a soft probability distribution over $\mathcal{U}$ using a Gaussian kernel \cite{bishop2006pattern}:
\begin{equation}
    p(u, i) = \exp\left( -\frac{\text{dist}(a_i, \text{center}(u))^2}{2\sigma^2} \right)
\end{equation}

This soft-binning allows images to partially cover adjacent units, accurately modeling inherent attribute ambiguity. To prevent unbounded spread and maintain computational efficiency, we apply a hard sparsity threshold: any bin assignment probability below 1\% is set to zero. Together with the natural decay of the Gaussian kernel, this bounds each image's effective coverage to a compact, semantically meaningful neighborhood of bins, reducing the soft-assignment matrix to a highly sparse structure. 

The representation of $a_i$ differs by attribute. For temporal attributes $a_i$ is a cyclical embedding, so the kernel wraps correctly across the 24-hour boundary. For geographic attributes $a_i$ is the (latitude, longitude) pair in degrees and $\mathcal{U}$ is a uniform $g \times g$ grid over $[-90^\circ, 90^\circ] \times [-180^\circ, 180^\circ]$, with $\text{dist}$ the Euclidean distance in degree space. $\sigma_{\text{geo}}$ is therefore expressed in degrees. This planar approximation neither wraps at the antimeridian nor corrects for the latitude-dependent contraction of longitude. Neither effect has a material impact on our benchmarks. No query's candidate set contains images on both sides of the $\pm 180^\circ$ boundary. Longitude distortion grows with latitude, and our splits are concentrated at low latitudes: images north of $60^\circ$ account for 0.4\%, 0.6\% and 4.6\% of the PixelProse, Incidents1M and SkyScript test splits respectively (Visual Genome is temporal-only), with no southern-polar images in any split. Only SkyScript carries a non-trivial high-latitude fraction, and its retrieval scores are near zero for reasons unrelated to binning (\extver{}). A geodesic kernel would nonetheless be the principled choice for globally uniform data. Note that the metadata channel of the diversity metric is computed on a different representation, unit 3D vectors on the sphere (\extver{}).

\subsection{Normalized Semantic Relevance ($\hat{R}$)}
A naive submodular coverage algorithm balances relevance and diversity. Let $r_i$ denote the raw semantic relevance score (e.g., the cosine similarity from a base VLM like CLIP or BLIP-2) between an image $i$ and the text query $q$. These raw scores $r_i$ are densely clustered and lack the scale required to compete with cumulative coverage sums. We define a locally normalized relevance $\hat{R}(i, q)$ over the candidate set $\mathcal{V}$:
\begin{equation}
    \hat{R}(i, q) = \frac{r_i - \min_{j \in \mathcal{V}}(r_j)}{\max_{j \in \mathcal{V}}(r_j) - \min_{j \in \mathcal{V}}(r_j) + \epsilon}
\end{equation}
This forces $\hat{R} \in [0, 1]$, ensuring the mathematical stability of the optimization trade-off.

\subsection{Query-Driven Bin Importance ($\Omega$)}

Standard probabilistic coverage \cite{xu2014efficient} attaches a non-negative weight $w_u$ to each Information Unit, but this weight is a static property of the unit, fixed prior to selection and uniform ($w_u = 1$) by default. Under query-independent weighting, an algorithm incentivized to cover every temporal bin will eventually select highly irrelevant images simply because they occupy an empty hour.

To prevent this, \mascot{} introduces dynamic, query-driven weights $\Omega(u, q)$, defined as the peak potential relevance within a given bin:
\begin{equation}
    \Omega(u, q) = \max_{j \in \mathcal{V}} \left( p(u, j) \cdot \hat{R}(j, q) \right)
\end{equation}
If a bin contains even one highly relevant image, $\Omega(u, q)$ approaches $1.0$, signaling the optimizer to cover it. If a bin contains only semantic noise, $\Omega(u, q) \to 0$; this naturally prevents the algorithm from wasting its retrieval budget on irrelevant outliers.

\subsection{The \mascot{} Objective}
By integrating these components, we formulate the Model-Aware Submodular Coverage for Composite-Attribute Text-to-Image Retrieval (\mascot{}). \mascot{} maximizes the following objective:
\begin{equation}
    \label{eq:masmf}
    f(\mathcal{S}) = (1-\lambda)\sum_{i \in \mathcal{S}} \hat{R}(i, q) + d \cdot \lambda \sum_{u \in \mathcal{U}} \Omega(u, q) \left(1 - \prod_{i \in \mathcal{S}} (1 - p(u, i))\right)
\end{equation}
where $\lambda \in [0, 1]$ is the user-defined parameter controlling the diversification intensity, and $d \in \{-1, +1\}$ is the task direction. For diversity-increase tasks, $d = +1$, rewarding broad coverage. For diversity-decrease tasks, $d = -1$, transforming the coverage term into a penalty that forces the model to tightly cluster selections within already-covered bins.

\subsection{Optimization and Relevance-Aligned Initialization}
\label{sec:penalty}
Optimizing the objective in Equation~\ref{eq:masmf} is NP-hard. We employ a standard greedy algorithm \cite{nemhauser1978analysis} that iteratively selects the image $i^*$ resulting in the maximum marginal gain $\Delta f(i | \mathcal{S}_{m-1})$ at each step $m$. 

The first selection deserves attention because it determines the top-ranked result. At $m=1$ the coverage state satisfies $P_{\text{covered}} = \mathbf{0}$, so the residual-capacity factor $(1 - P_{\text{covered},u})$ is at its maximum and $\Delta_{\text{cov}}(i) = \sum_u \Omega(u,q)\,p(u,i)$ is the largest value the coverage term ever takes. It is non-zero for every candidate and varies across them, so the first selection is governed jointly by $\hat{R}$ and $\Delta_{\text{cov}}$.

The two terms are correlated by construction: $\Omega$ is built from $\hat{R}$, so the most relevant image necessarily occupies a high-$\Omega$ bin and receives a high $\Delta_{\text{cov}}$. Under $d = +1$ they reinforce and the top-ranked result is largely retained. Under $d = -1$ they oppose, and the most relevant image is also the most penalized. This is the mechanism behind the top-1 displacement quantified in \extver{}: on the three PixelProse decrease tasks, \mascot{}'s first selection matches the base retriever's on 84.4\%, 76.0\% and 65.8\% of queries respectively. Displacement is an expected consequence of concentration rather than an artifact, and the resulting top-1 accuracy stays well above what manifold repulsion retains.

\section{Algorithm}

Optimizing the \mascot{} objective (Equation~\ref{eq:masmf}) exactly is NP-hard. The base probabilistic coverage function $\mathrm{cov}(\mathcal{S})$ is monotone and submodular \cite{xu2014efficient}.

\textbf{Increase mode.} Here $f^{+}(\mathcal{S}) = (1-\lambda)\hat{R}(\mathcal{S}) + \lambda\,\mathrm{cov}(\mathcal{S})$ is the sum of a modular relevance term and a monotone submodular coverage term, and is therefore itself monotone submodular. Greedy selection under a cardinality constraint attains the classical $(1-1/e)$ approximation guarantee \cite{nemhauser1978analysis}.

\textbf{Decrease mode.} Here $f^{-}(\mathcal{S}) = (1-\lambda)\hat{R}(\mathcal{S}) - \lambda\,\mathrm{cov}(\mathcal{S})$. Since $\mathrm{cov}$ is monotone submodular, $-\mathrm{cov}$ is supermodular and $f^{-}$ is a difference-of-submodular objective \cite{iyer2012algorithms}, for which the $(1-1/e)$ bound does not apply. We therefore treat greedy selection in decrease mode as a principled heuristic rather than an approximation algorithm, and validate it empirically through Recall@K, the Preference Reflection Score, and the early-rank analysis of Section~\ref{sec:results}. In practice the cardinality ratio $K/N = 0.1$ tightly bounds the feasible search space.

Algorithm~\ref{alg:mascot} details the MASCOT subset selection process. By maintaining a state vector of the cumulative probability coverage ($P_{\text{covered}}$), we efficiently compute the marginal coverage gain $\Delta_{\text{cov}}$ at each step.

\begin{algorithm}[t]
   \caption{\mascot{} Greedy Subset Selection}
   \label{alg:mascot}
\begin{algorithmic}[1]
   \STATE {\bfseries Input:} Candidate set $\mathcal{V}$, VLM scores $r$, Metadata attributes $\mathcal{A}$, subset size $K$, intensity $\lambda$, direction $d \in \{-1, 1\}$
   \STATE {\bfseries Output:} Selected subset $\mathcal{S}$
   
   \STATE \texttt{\% 1. Precomputation Phase}
   \STATE $\hat{R} \leftarrow \text{Normalize}(r)$ over $\mathcal{V}$
   \STATE $p \leftarrow \text{GaussianSoftBinning}(\mathcal{A}, \mathcal{U})$
   \STATE $\Omega \leftarrow \text{ComputeBinImportance}(p, \hat{R})$
   
   \STATE \texttt{\% 2. Initialization Phase}
   \STATE $\mathcal{S} \leftarrow \emptyset$
   \STATE $P_{\text{covered}} \in \mathbb{R}^{|\mathcal{U}|} \leftarrow \mathbf{0}$
   
   \STATE \texttt{\% 3. Greedy Selection Loop}
   \FOR{$m=1$ {\bfseries to} $K$}
       \FOR{{\bfseries each} $i \in \mathcal{V} \setminus \mathcal{S}$}
           \STATE \texttt{\% Calculate Marginal Coverage Gain}
           \STATE $\Delta_{\text{cov}}(i) = \sum_{u \in \mathcal{U}} \Omega(u) \cdot (1 - P_{\text{covered}, u}) \cdot p(u, i)$
           
           \STATE \texttt{\% Apply Directional Logic}
           \IF{$d == -1$ (Diversity Decrease Task)}
               \STATE $\text{Gain}(i) = (1-\lambda)\hat{R}_i - \lambda \Delta_{\text{cov}}(i)$
           \ELSE
               \STATE $\text{Gain}(i) = (1-\lambda)\hat{R}_i + \lambda \Delta_{\text{cov}}(i)$
           \ENDIF
       \ENDFOR
       
       \STATE $i^* = \arg\max_{i} \text{Gain}(i)$
       
       \STATE $P_{\text{covered}} \leftarrow 1 - (1 - P_{\text{covered}}) \odot (1 - p_{i^*})$
       \STATE $\mathcal{S} \leftarrow \mathcal{S} \cup \{i^*\}$
   \ENDFOR
   
   \STATE {\bfseries Return} $\mathcal{S}$
\end{algorithmic}
\end{algorithm}

As demonstrated, as a specific Information Unit $u$ becomes highly covered ($P_{\text{covered}, u} \to 1$), the marginal gain for selecting any subsequent image that maps to $u$ effectively vanishes. This forces the model to dynamically shift its focus to uncovered bins, driving the robust attribute diversification characteristic of the MASCOT framework.

\subsection{Theoretical Analysis: Saturation vs. Spatial Repulsion}
\label{sec:theory}

The primary theoretical distinction between \mascot{} and manifold-based approaches such as MS-DPP lies in their saturation behavior, which dictates how they handle diversity decrease constraints.

In MS-DPP, diversity is enforced by maximizing the determinant of the kernel, which corresponds to the geometric volume of the parallelepiped spanned by the selected image vectors in the SPD tangent space. This spatial repulsion is continuous and global. When the task demands decreasing diversity, MS-DPP attempts to minimize this volume. However, because the repulsion is continuous, the model is penalized for selecting semantically relevant images that naturally occupy distinct visual manifolds, producing the recall degradation we quantify in Section~\ref{sec:results}.

In contrast, \mascot{} is governed by the submodular principle of diminishing marginal returns, specifically tied to the state vector $P_{\text{covered}}$. This fundamentally alters the optimization dynamics based on the task direction $d$:
\begin{itemize}
    \item \textbf{Diversity Increase ($d = +1$):} The marginal coverage gain $\Delta_{\text{cov}}$ acts as a reward. When a bin $u$ is empty, $\Delta_{\text{cov}}$ is high. As the bin saturates ($P_{\text{covered}, u} \to 1$), the reward for selecting images in that bin vanishes, gracefully forcing the algorithm to explore the uncovered bins.
    \item \textbf{Diversity Decrease ($d = -1$):} The coverage gain $\Delta_{\text{cov}}$ is inverted into a penalty. Selecting an image from a new, empty bin yields a high penalty. Crucially, as a bin saturates ($P_{\text{covered},u} \to 1$), the penalty on images mapping to it decays toward zero, and the selection increasingly defaults to normalized semantic relevance $\hat{R}_i$. Because soft-binning spreads each image across several bins, this decay is partial rather than complete: an image is penalized in proportion to the residual capacity of every bin it touches.

\end{itemize}

Manifold approaches treat decreasing diversity as a spatial compression problem, where minimizing the geometric volume forces out distant semantic matches with no lower bound on the penalty. Coverage instead bounds the penalty by the residual capacity of the bins an image occupies, so selections drawn from an already-concentrated region are penalized weakly and the objective reverts toward semantic relevance. This bounds rather than eliminates the loss of early-rank relevance: \mascot{} does displace the top-ranked result under decrease constraints, but recovers quickly with rank, reaching R@10 = 0.9410 on \texttt{PP\_geo\_hour} where manifold repulsion remains at 0.4931 (Section~\ref{sec:results}).

\section{Experimental Setup}
\subsection{Dataset and Tasks}
We evaluate \mascot{} primarily on PixelProse \cite{singla2024pixelsproselargedataset}, a large-scale multimodal dataset of over 16M images with dense synthetic captions. After filtering for images whose EXIF metadata contains both valid GPS coordinates and a capture timestamp, and discarding entries whose source URLs no longer resolve, we retain 996 images (\extver{}). These are substantially fewer than the 25{,}151 reported by \citet{sogi2025msdppsmultisourcedeterminantalpoint}, a gap we attribute to updates in the PixelProse release and to continued decay of source image URLs; because the same 996 images are used for \mascot{} and all baselines, relative comparisons remain valid. We additionally report results on Visual Genome, Incidents1M, and SkyScript in \extver{}. Following the CDR-CA evaluation protocol, we assess performance on geographic coordinates (Latitude/Longitude), temporal metadata (Hour of day) and their combination. We define two distinct task directions:
\begin{itemize}
    \item \textbf{Diversity (Increase):} Maximizing metadata diversity while retaining semantic relevance (e.g., fetching visually relevant images from geographically diverse locations).
    \item \textbf{Diversity (Decrease):} Minimizing metadata diversity to strictly bound the retrieval space (e.g., forcing the VLM to fetch relevant images from a specific, narrow time window).
\end{itemize}

\subsection{Metrics}
We measure semantic retention using \textbf{Recall@10 (R@10)}, using BLIP-2 \cite{li2023blip2bootstrappinglanguageimagepretraining} as the base dense retriever to generate a candidate set of size $N=200$, from which subsets of $K=20$ are selected.\footnote{\citet{sogi2025msdppsmultisourcedeterminantalpoint} report N@10 on PixelProse; we report R@10, which is better suited to our single-ground-truth setting where each caption matches exactly one image. Absolute values are therefore not directly comparable to theirs, though all methods here are evaluated identically.} For attribute distribution we report the \textbf{Diversity Metric (DM)} of \citet{sogi2025msdppsmultisourcedeterminantalpoint}, a normalized Vendi Score \cite{friedman2023vendiscorediversityevaluation} combining an appearance and a
metadata channel. The full definition is given in \extver{}. DM is reported so that higher is better in both directions: on decrease tasks the metadata channel is inverted
before combination, so a higher DM indicates stronger metadata concentration. Unless stated as \emph{raw} diversity, all reported DM values use this higher-is-better transformation.
Unless otherwise stated, geographic tasks use a $20 \times 20$ grid ($|\mathcal{U}_{\text{geo}}| = 400$) and temporal tasks 24 hourly bins. Per-task values of $\lambda$, $\sigma_{\text{geo}}$ and $\sigma_{\text{time}}$, selected by validation grid search, are reported in full in \extver{}.

\subsection{Baselines}
We benchmark \mascot{} against the unconstrained dense retriever (BLIP-2) \cite{li2023blip2bootstrappinglanguageimagepretraining}, k-DPP \cite{10.5555/3104482.3104632}, clustering \cite{boteanu2017pseudo}, MMR \cite{10.1145/290941.291025}, standard Probabilistic Coverage (Adapted from \cite{xu2014efficient}), and state-of-the-art manifold-based approaches: MS-DPP and its advanced variants \cite{sogi2025msdppsmultisourcedeterminantalpoint}.

We evaluate three MS-DPP variants: the base method, one with tangent normalization (TN), and one that additionally normalizes the unified tangent vector (TVMS). Implementation details are in \extver{}.

We note that our Prob-Coverage ablation is itself an adaptation of the submodular coverage framework of \citet{xu2014efficient}, and thus serves as a representative submodular-retrieval baseline in addition to functioning as an ablation of \mascot{}.

We defer the full sensitivity sweeps to \extver{}.

\section{Results and Discussion}
\label{sec:results}
\subsection{Decrease Tasks}

The most critical test of a controllable retriever's boundary enforcement is the strict diversity-decrease task: tightly decreasing metadata variance while maintaining high semantic relevance. Table \ref{tab:decrease} presents these results, where the goal is to maximize the normalized Diversity Metric ($\uparrow$) without sacrificing early-rank retrieval accuracy (R@10). 

\begin{table*}[!t]
\begin{center}
\begin{small}
\begin{tabular}{lccc}
\toprule
& \textbf{PP\_geo\_hour} & \textbf{PP\_hour} & \textbf{PP\_geo} \\
\textbf{Method} & \textbf{HM$\uparrow$(R@10$\uparrow$, DM$\uparrow$)} & \textbf{HM$\uparrow$(R@10$\uparrow$, DM$\uparrow$)} & \textbf{HM$\uparrow$(R@10$\uparrow$, DM$\uparrow$)} \\
\midrule
BLIP-2 (Base)   & 0.2831 (0.9737, 0.1656) & 0.4593 (0.9737, 0.3006) & 0.5055 (0.9737, 0.3414) \\
\midrule
Clustering      & 0.3205 (0.7240, 0.2058) & 0.4708 (0.7189, 0.3500) & 0.5089 (0.7014, 0.3993) \\
MMR             & 0.2448 (0.9724, 0.1401) & 0.4400 (0.9762, 0.2840) & 0.4683 (0.9711, 0.3085) \\
k-DPP           & 0.2741 (0.8959, 0.1618) & 0.4614 (0.9511, 0.3046) & 0.5035 (0.9724, 0.3397) \\
\midrule
MS-DPP          & 0.3850 (0.4931, 0.3158) & 0.5503 (0.7654, 0.4296) & 0.6270 (0.7704, 0.5286) \\
MS-DPP+TN       & 0.3582 (0.8934, 0.2240) & 0.5342 (0.8821, 0.3831) & 0.6337 (0.7829, 0.5322) \\
MS-DPP+TN+TVMS  & 0.3363 (0.9021, 0.2066) & 0.5205 (0.9072, 0.3649) & 0.6260 (0.7917, 0.5176) \\
\midrule
\textbf{Ours (Prob-Coverage/Both Ablated)} & \underline{0.4066} (0.7302, 0.2817) & \underline{0.6186} (0.8143, 0.4988) & \textbf{0.6409} (0.8670, 0.5084) \\
\textbf{Ours (w/o Normalization)} & 0.2616 (0.2873, 0.2401) & 0.3487 (0.2974, 0.4215) & 0.5026 (0.5947, 0.4352) \\
\textbf{Ours (Uniform Binning)} & \textbf{0.4240} (0.8118, 0.2870) & \textbf{0.6197} (0.8381, 0.4915) & \underline{0.6397} (0.8695, 0.5060) \\
\textbf{Ours (\mascot{})} & 0.3135 (0.9410, 0.1881) & 0.5121 (0.9059, 0.3570) & 0.5478 (0.8105, 0.4136) \\
\bottomrule
\end{tabular}
\end{small}
\end{center}
\caption{Comparative results on decreasing attribute diversity task. The best and second-best among diversification methods are highlighted in bold and underlined, respectively.}
\label{tab:decrease}
\end{table*}

The empirical data validates our theoretical analysis of manifold vulnerabilities. When forced to tightly bound the retrieval space across both geography and time (\texttt{PP\_geo\_hour}), the continuous repulsion mechanism of the base MS-DPP \cite{sogi2025msdppsmultisourcedeterminantalpoint} discards semantic matches, causing R@10 to collapse from 0.9737 to 0.4931. While the TVMS variant rescues some recall (0.9021), it struggles to meaningfully improve the Diversity Metric over the baseline (0.2066 vs 0.1656).

By breaking our submodular formulation into its ablations, the mechanics of MASCOT become clear. The naive \textbf{Prob-Coverage} base variant (which lacks both normalization and query-driven binning) behaves blindly; it heavily sacrifices early-rank recall (0.7302) to forcefully inflate the harmonic mean. Removing the local relevance normalization (\textbf{w/o Normalization}) causes severe failure, as the unscaled VLM cosine similarities are completely overwhelmed by the cumulative coverage penalties, dropping R@10 to 0.2873. 

Among methods that genuinely compress diversity, i.e., that push DM above the unconstrained baseline, \mascot{} attains the highest R@10 on \texttt{PP\_geo\_hour} (0.9410 at DM = 0.1881). Heuristics such as MMR and k-DPP retain high recall only by failing to compress: their DM (0.1401 and 0.1618) falls below the baseline's 0.1656, so they are not performing the decrease task at all.

The ablations, however, complicate a simple reading. \textbf{Uniform Binning} (\mascot{} without query-driven weighting) attains a higher harmonic mean on all three decrease tasks, compressing harder in every case (DM 0.2870, 0.4915, 0.5060 against 0.1881, 0.3570, 0.4136). The contribution of $\Omega$ is therefore not aggregate score but the shape of the recall curve. \mascot{} recovers to R@10 = 0.9410 on the composite task where Uniform Binning plateaus at 0.8118. On \texttt{PP\_geo}, where the metadata admits no composite structure, Uniform Binning is better at every rank and the query-driven weighting provides no benefit.

\subsection{Increase Tasks}

Table \ref{tab:increase} evaluates the frameworks on standard diversity maximization tasks, where the goal is to increase attribute variance. 

\begin{table*}[!t]
\begin{center}
\begin{small}
\begin{tabular}{lccc}
\toprule
& \textbf{PP\_geo\_hour} & \textbf{PP\_hour} & \textbf{PP\_geo} \\
\textbf{Method} & \textbf{HM$\uparrow$(R@10$\uparrow$, DM$\uparrow$)} & \textbf{HM$\uparrow$(R@10$\uparrow$, DM$\uparrow$)} & \textbf{HM$\uparrow$(R@10$\uparrow$, DM$\uparrow$)} \\
\midrule
BLIP-2 (Base)   & 0.9430 (0.9737, 0.9142) & 0.9192 (0.9737, 0.8706) & 0.9039 (0.9737, 0.8435) \\
\midrule
Clustering      & 0.9005 (0.8657, 0.9381) & 0.8838 (0.8758, 0.8920) & 0.8759 (0.8620, 0.8904) \\
MMR             & 0.9496 (0.9636, 0.9359) & \underline{0.9285} (0.9737, 0.8873) & 0.8919 (0.8670, 0.9182) \\
k-DPP           & \textbf{0.9533} (0.9686, 0.9384) & \textbf{0.9331} (0.9649, 0.9034) & \textbf{0.9291} (0.9674, 0.8938) \\
\midrule
MS-DPP          & 0.9487 (0.9724, 0.9262) & 0.9277 (0.9410, 0.9147) & 0.9127 (0.8984, 0.9274) \\
MS-DPP+TN       & 0.9424 (0.9435, 0.9413) & 0.9197 (0.9310, 0.9087) & 0.8985 (0.8758, 0.9223) \\
MS-DPP+TN+TVMS  & \underline{0.9504} (0.9711, 0.9306) & 0.9248 (0.9749, 0.8796) & \underline{0.9236} (0.9586, 0.8910) \\
\midrule
\textbf{Ours (Prob-Coverage/Both Ablated)} & 0.9401 (0.9473, 0.9330) & 0.9011 (0.8833, 0.9195) & 0.9154 (0.9009, 0.9304) \\
\textbf{Ours (w/o Normalization)} & 0.9119 (0.8871, 0.9381) & 0.9105 (0.9072, 0.9139) & 0.8920 (0.9398, 0.8488) \\
\textbf{Ours (Uniform Binning)} & 0.9224 (0.9072, 0.9381) & 0.9066 (0.8946, 0.9189) & 0.9074 (0.8821, 0.9343) \\
\textbf{Ours (\mascot{})} & 0.8863 (0.8356, 0.9435) & 0.8951 (0.8921, 0.8982) & 0.9084 (0.9109, 0.9059) \\
\bottomrule
\end{tabular}
\end{small}
\end{center}
\caption{Comparative results on diversity increasing tasks. The best and second-best among diversification methods are highlighted in bold and underlined, respectively.}
\label{tab:increase}
\end{table*}

Because spatial repulsion naturally pushes items apart on the manifold, MS-DPP variations easily achieve high diversity while retaining strong retrieval accuracy. However, \mascot{} remains highly competitive. By relying on the principle of diminishing marginal returns, \mascot{} naturally explores uncovered bins, achieving the highest absolute Diversity Metric (0.9435) on the composite \texttt{PP\_geo\_hour} task, though with a corresponding drop in recall. Spatial repulsion is optimized for one-directional exploration, and MS-DPP variants exploit this to lead on several increase settings. \mascot{} remains competitive here while avoiding the reversal penalty: MS-DPP's advantage on increase does not transfer to decrease, where its recall falls to 0.4931 on the composite task against \mascot{}'s 0.9410. The coverage formulation trades a modest amount of increase-direction performance for behavior that does not degenerate when the constraint is inverted.

\subsection{Early-Rank Semantic Integrity}

While standard retrieval metrics like R@10 capture general subset relevance, production search systems heavily prioritize early-rank semantic integrity (the top 1 to 5 results). Spatial repulsion models continuously penalize the geometric volume on the manifold \cite{Kulesza_2012}. We therefore examine the full recall curve rather than R@10 alone, asking not only how much relevance each method retains but at which ranks it is lost.

\begin{figure*}[!t]
    \centering
    \begin{subfigure}{0.3\textwidth}
        \includegraphics[width=\linewidth]{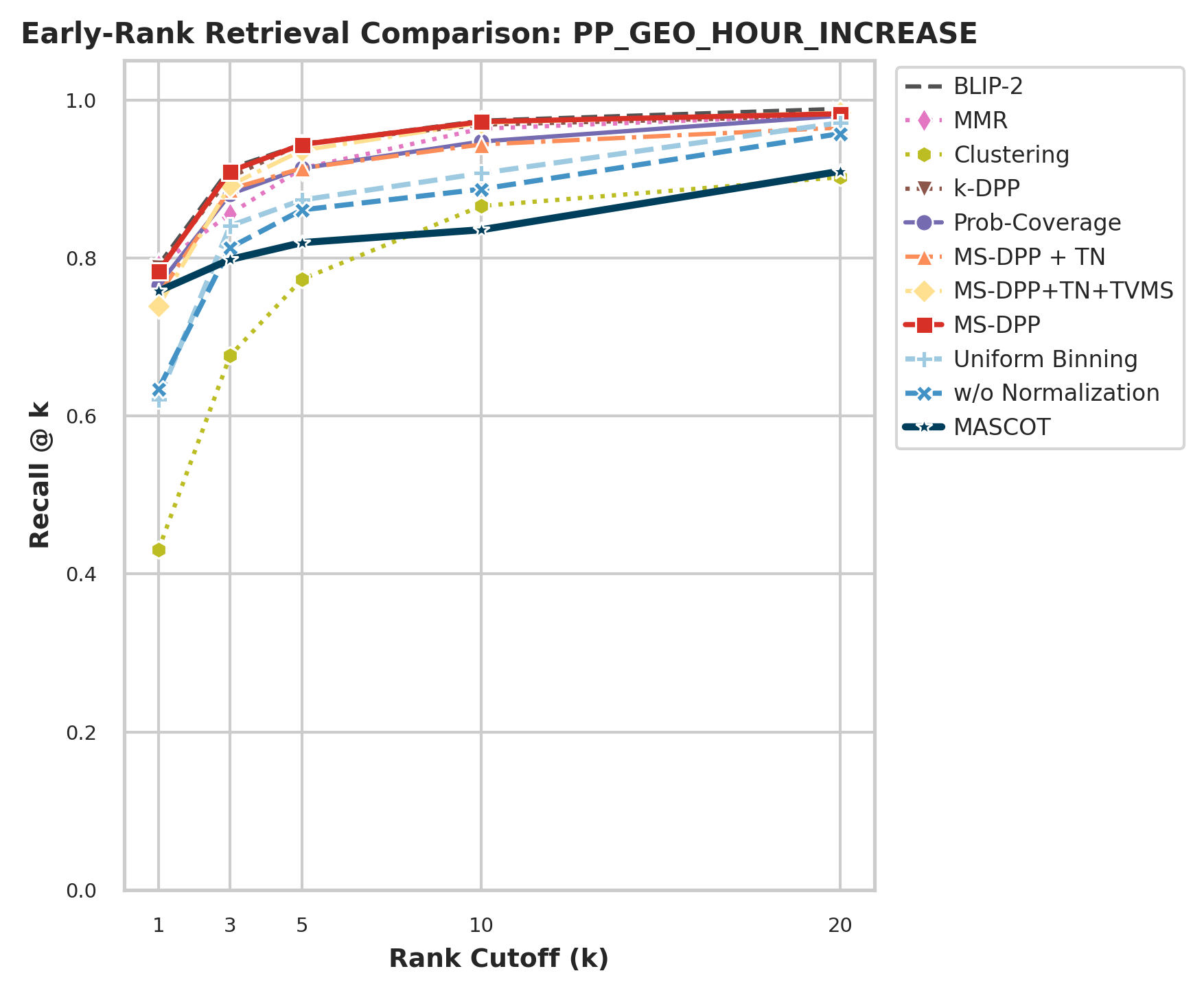}
        \caption{Geo + Hour (Increase)}
    \end{subfigure}\hfill
    \begin{subfigure}{0.3\textwidth}
        \includegraphics[width=\linewidth]{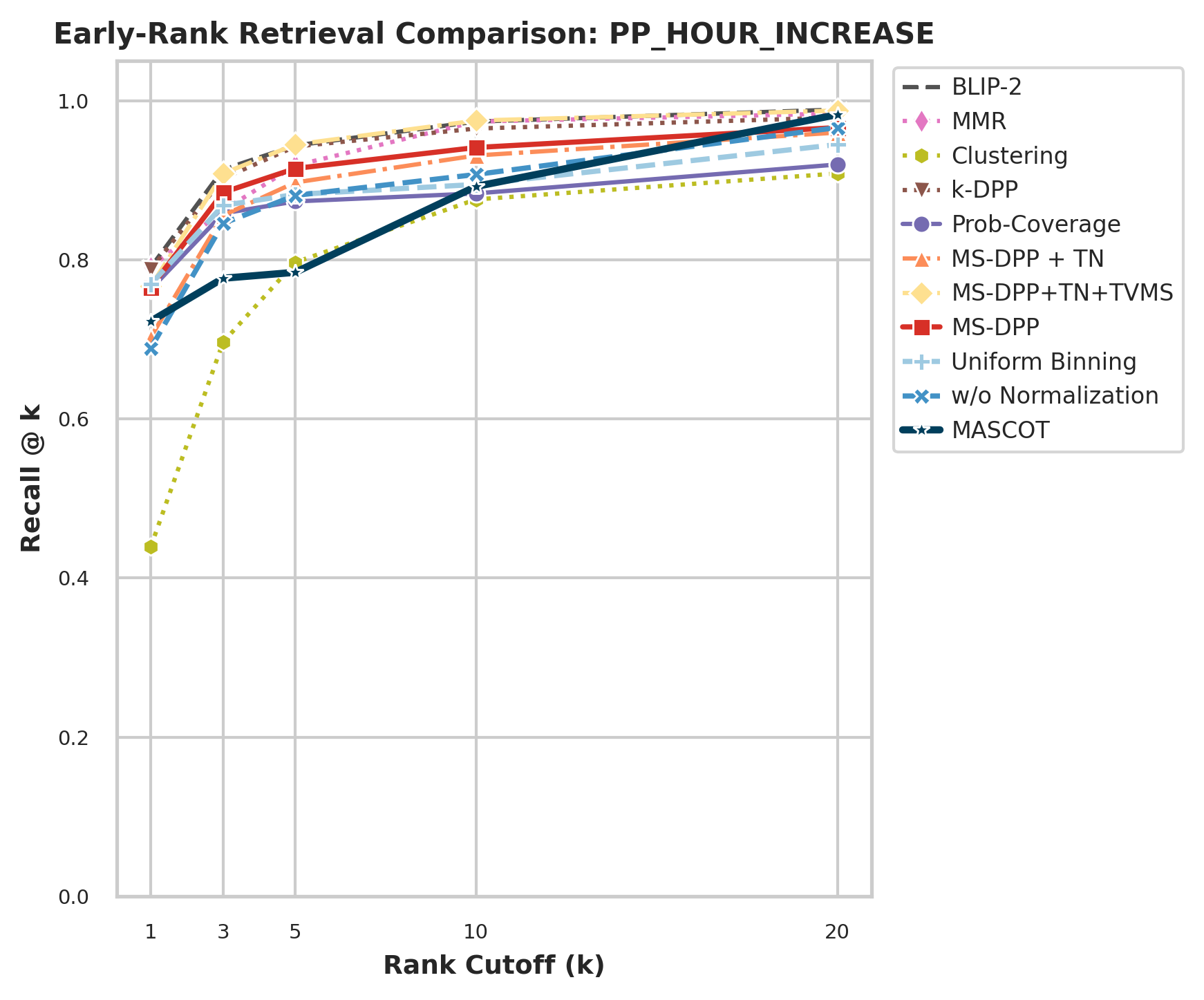}
        \caption{Hour (Increase)}
    \end{subfigure}\hfill
    \begin{subfigure}{0.3\textwidth}
        \includegraphics[width=\linewidth]{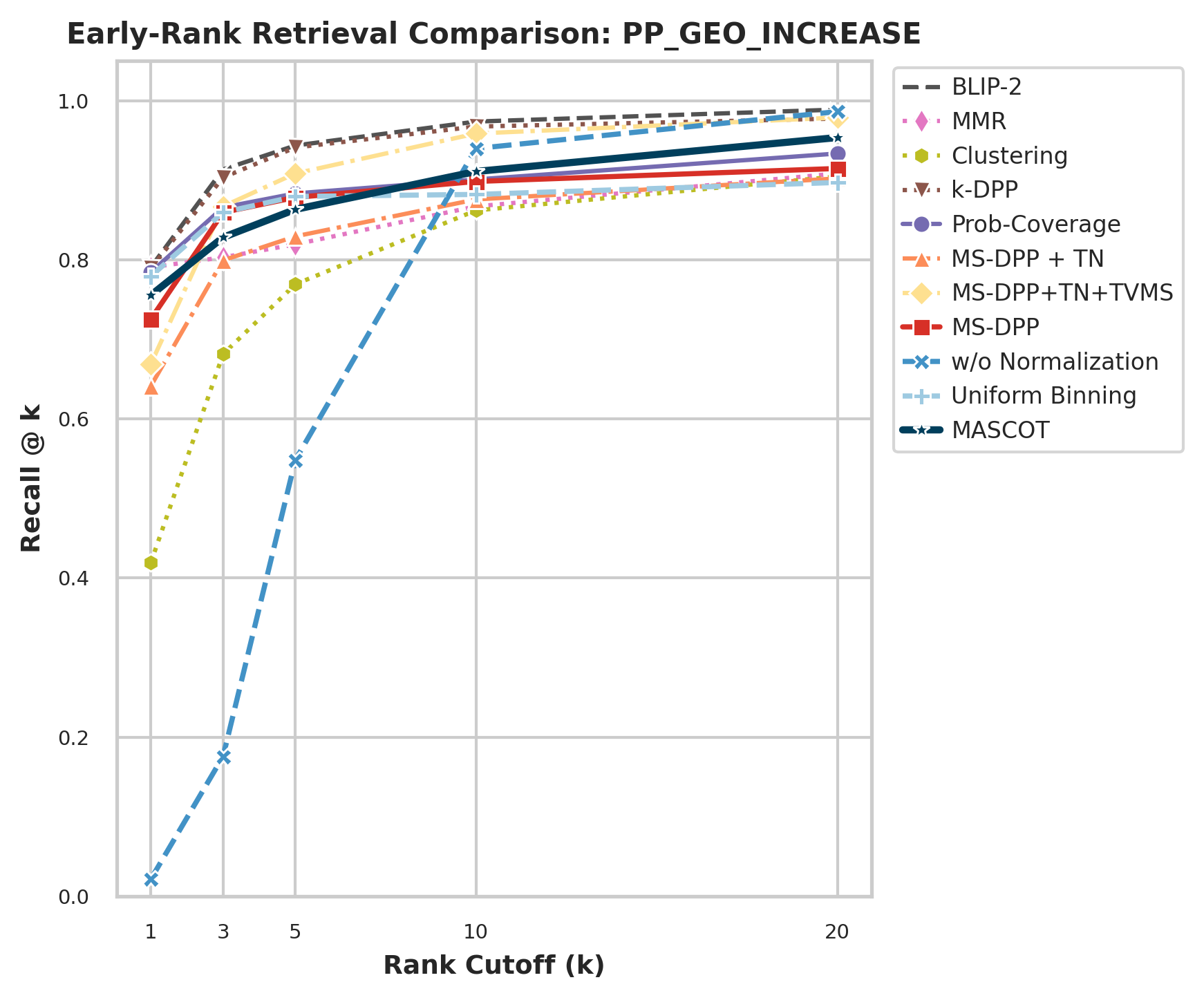}
        \caption{Geo (Increase)}
    \end{subfigure}
    
    \vspace{1mm} 
    
    \begin{subfigure}{0.3\textwidth}
        \includegraphics[width=\linewidth]{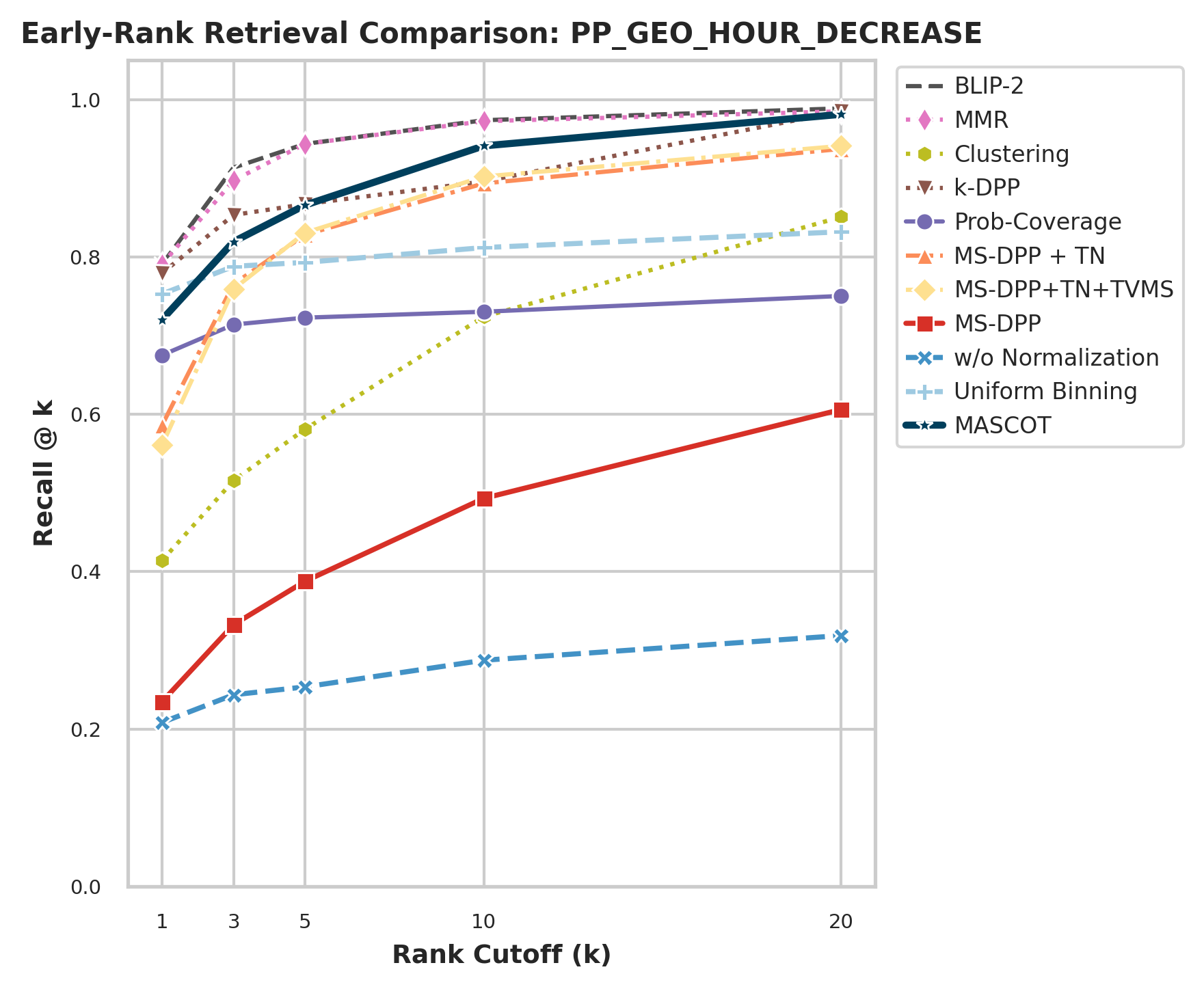}
        \caption{Geo + Hour (Decrease)}
    \end{subfigure}\hfill
    \begin{subfigure}{0.3\textwidth}
        \includegraphics[width=\linewidth]{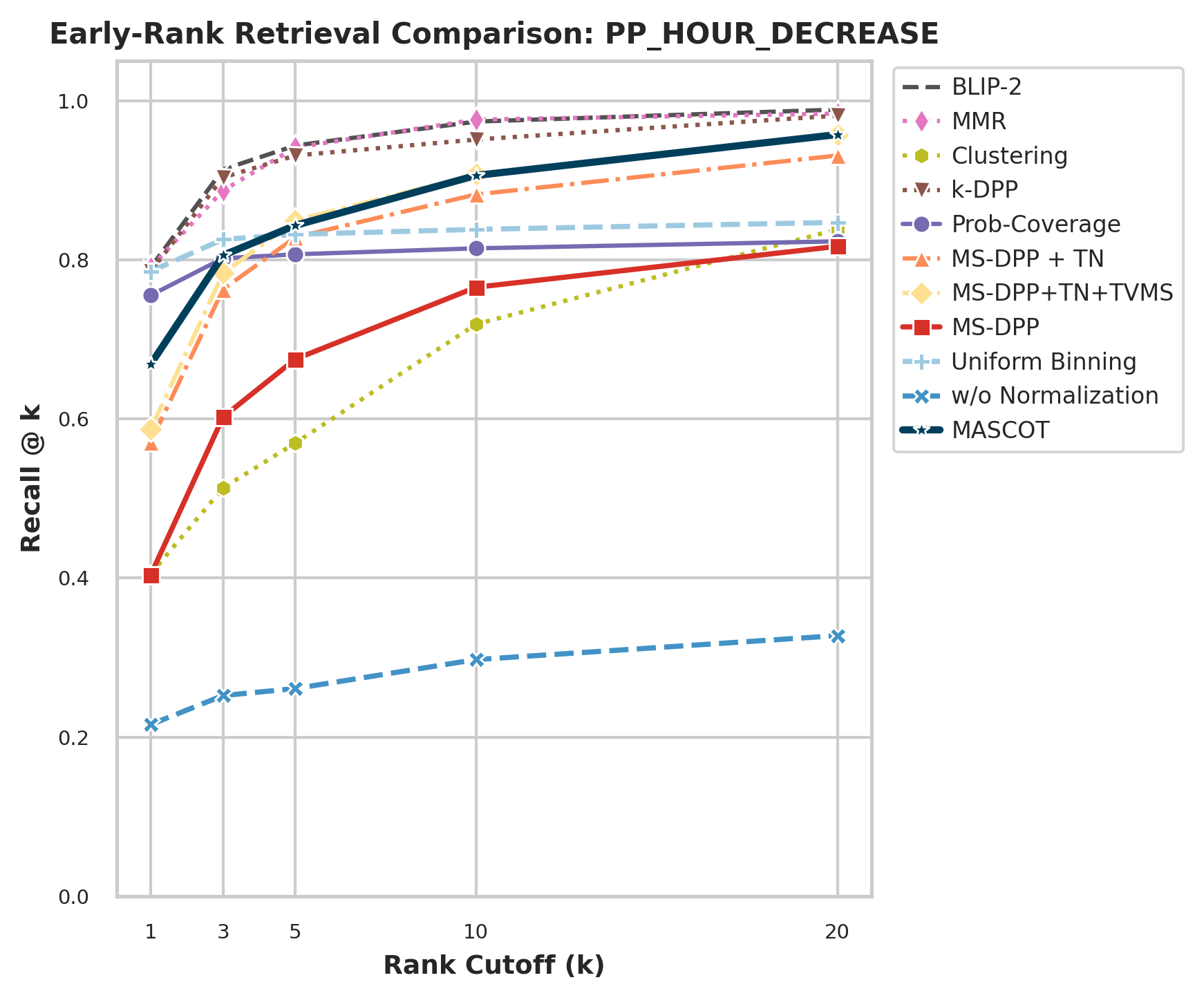}
        \caption{Hour (Decrease)}
    \end{subfigure}\hfill
    \begin{subfigure}{0.3\textwidth}
        \includegraphics[width=\linewidth]{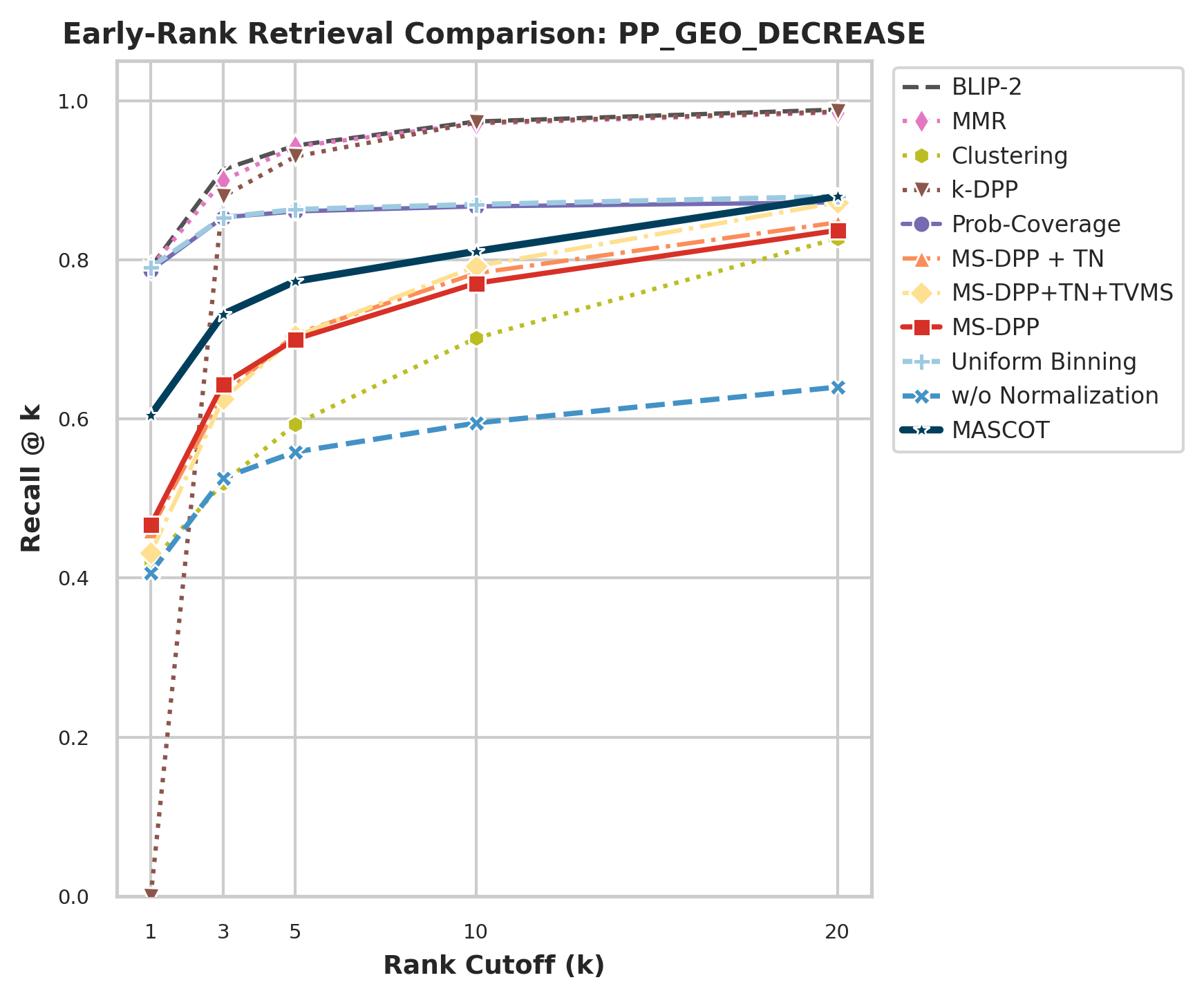}
        \caption{Geo (Decrease)}
    \end{subfigure}
    \vspace{-2mm}
    \Description{Six Recall@K plots for K from 1 to 20 on three PixelProse tasks in both directions, comparing MASCOT against MS-DPP variants and baselines.}
    \caption{Recall@K curves ($K \in [1, 20]$) across all datasets and tasks. \textbf{Top Row (Increase):} Continuous manifold repulsion remains stable when expanding the retrieval space, with most models closely tracking the BLIP-2 baseline. \textbf{Bottom Row (Decrease):} When forced to tightly bound attributes, manifold models (MS-DPP variants) and older algorithms (k-DPP, Clustering) suffer severe early-rank degradation on the PixelProse tasks, dropping sharply at rank~1. \mascot{} degrades more gradually and recovers faster with rank, though it does not preserve the top-ranked result.
}
    \label{fig:recall_curves}
\end{figure*}

Figure \ref{fig:recall_curves} provides a comprehensive visual validation of this phenomenon across the complete retrieval window ($K=1$ to $20$). During diversity increase tasks (Figure \ref{fig:recall_curves}, Top Row), where spatial repulsion aligns naturally with expanding the retrieval space, the recall curves for the manifold models remain tightly clustered near the unconstrained BLIP-2 baseline. 

However, the moment the task shifts to decreasing diversity (Figure \ref{fig:recall_curves}, Bottom Row), the fragility of continuous manifold optimization is visually exposed. Across all three PixelProse tasks, the base MS-DPP framework degrades sharply at rank 1, dropping to 0.2346 on the composite \texttt{PP\_geo\_hour} task and recovering slowly as $K$ increases. Even the highly augmented MS-DPP+TN+TVMS variant routinely fails to protect the top matches. Furthermore, older baselines exhibit severe instability, highlighted by the k-DPP curve flatlining at zero for the initial rank on the geographic constraint task.

\mascot{} degrades far more gracefully, though it does not preserve the top-ranked result. On \texttt{PP\_geo\_hour} decrease it retains R@1 = 0.7202 against the unconstrained baseline's 0.7905 and MS-DPP's 0.2346, and recovers to 0.9410 by rank~10. Averaged over the three PixelProse decrease tasks it holds R@1 = 0.665 against MS-DPP's 0.369. The distinguishing property is not rank-1 protection but the shape of the curve: coverage-based selection loses a bounded amount of relevance at the head and recovers it quickly, whereas manifold repulsion loses more and recovers slowly.

The comparison against our own ablation sharpens this. Uniform Binning attains a higher R@1 than \mascot{} on all three decrease tasks (0.7528, 0.7854, 0.7905), but its curve is nearly flat: on \texttt{PP\_geo\_hour} it reaches only 0.8118 by rank~10, where \mascot{} reaches 0.9410. Query-driven weighting thus costs early-rank accuracy and buys mid-rank recovery under composite constraints. On the single-attribute \texttt{PP\_geo} task it buys nothing: Uniform Binning is equal or better at every rank.

This saturation behavior has no direct analogue in manifold-based methods. Because DPP-based repulsion evaluates the volume of the selected parallelepiped, a candidate's penalty shrinks only as it becomes similar to what is already selected, with no floor below which coverage of a region exempts further selections from it. \mascot{}'s penalty is instead bounded by the residual capacity of the bins an image occupies, which is why its recall recovers quickly with rank rather than remaining depressed across the whole curve.

Qualitative visualizations of retrieved sets, geographic heatmaps and temporal histograms are provided in \extver{}.



\subsection{Operating Regimes and Limitations}
\label{sec:scope}
Three further analyses appear in \extver{}. \mascot{} attains perfect Preference Reflection Score on four of six task-directions without manifold normalization, trailing tangent-normalized MS-DPP on the two composite and geographic decrease settings (\extver{}). Its behavior depends jointly on $\lambda$, the soft-binning bandwidths and the grid resolution, and the configurations reported above sit one grid step above an abrupt collapse along most axes, a limitation we quantify and give a detection procedure for in \extver{}. Finally, the advantage over manifold repulsion is not uniform across datasets: it requires capture-accurate metadata and a candidate pool large enough for repulsion to overreach, conditions that fail on \texttt{VG\_hour} and \texttt{I1M\_geo} (\extver{}). The objective itself is attribute-agnostic and extends to learned cluster metadata without modification (\extver{}).

\section{Conclusion}

We identified a failure mode in state-of-the-art controllable retrieval: when continuous manifold-based methods such as MS-DPP \cite{sogi2025msdppsmultisourcedeterminantalpoint} are forced to \emph{decrease} attribute diversity, their spatial repulsion penalizes geometric volume and discards semantically relevant images, collapsing early-rank recall. The effect is most severe under composite constraints, where suppression is demanded along several attribute axes at once.

To address this, we introduced \mascot{}, which reframes diversity not as spatial distance but as dynamic coverage over discrete, query-weighted Information Units, grounded in submodular optimization \cite{nemhauser1978analysis,xu2014efficient}. Because the coverage penalty is bounded by the residual capacity of each bin rather than growing without limit, the relevance lost at early ranks is bounded and recovered quickly: on \texttt{PP\_geo\_hour} decrease \mascot{} retains R@1 = 0.7202 and R@10 = 0.9410, against MS-DPP's 0.2346 and 0.4931. We do not claim uniform superiority. On aggregate diversity--relevance scores our own Uniform Binning ablation attains higher harmonic means on all three decrease tasks, and on the single-attribute \texttt{PP\_geo} task it is better at every rank; on datasets with small candidate pools or coarse metadata the manifold approach can lead. The contribution of query-driven weighting is specific to composite constraints, where it trades early-rank accuracy for substantially faster recovery with rank. \mascot{}'s contribution is a coverage formulation that handles the full increase--decrease spectrum without the sharp early-rank degradation manifold methods suffer when the constraint is inverted, together with a characterization (\extver{} and Section~\ref{sec:scope}) of the operating regimes and hyperparameter margins in which it applies. Reducing the greedy re-ranker's latency and extending the framework to learned or continuous metadata are natural directions for future work.

\section*{Generative AI Disclosure}
In accordance with the ACM Policy on Authorship, the authors disclose the use of generative AI tools limited to improving the grammar and phrasing of author-written text and assisting with code implementation and \LaTeX{} formatting. These tools were not used to generate research ideas, experimental results, or citations. The authors have verified the references cited in this work and assume full responsibility for the content of this work.

\clearpage
\bibliographystyle{ACM-Reference-Format}
\balance
\bibliography{sample-base}
\clearpage
\appendix

\section{Text-to-Image Retrieval}
\label{sec:appendix_t2ir}

This appendix expands the overview in Section~\ref{sec:related} (Related Work).
Text-to-image retrieval aligns visual and textual modalities in a shared space.
Foundational works approached this through visual-semantic embeddings
\cite{10.5555/2999792.2999849,faghri2018vseimprovingvisualsemanticembeddings},
and learning objectives subsequently divided into two broad paradigms.
\emph{Dual-encoder} models
\cite{radford2021learningtransferablevisualmodels,jia2021scalingvisualvisionlanguagerepresentation,yuan2021florencenewfoundationmodel}
encode each modality independently and align the two spaces with a contrastive
or ranking objective, so that retrieval reduces to nearest-neighbour search over
embeddings that can be precomputed offline. \emph{Fusion-based} models
\cite{NEURIPS2021_50525975,chen2020uniteruniversalimagetextrepresentation}
instead encode an image--text pair jointly and predict a matching score,
treating relevance as a classification problem over the fused representation.
This yields finer-grained alignment but requires a forward pass per candidate
pair, so the two objectives are frequently combined: a contrastive stage
retrieves a shortlist that a matching head then re-scores
\cite{NEURIPS2021_50525975}.

To capture localized correspondence between regions and words, models leverage
cross-attention mechanisms such as SCAN
\cite{lee2018stackedcrossattentionimagetext}, and transformer-based fusion
architectures \cite{alayrac2022flamingovisuallanguagemodel,yu2022cocacontrastivecaptionersimagetext}.
Single-stream early-fusion designs including UNITER
\cite{chen2020uniteruniversalimagetextrepresentation}, OSCAR
\cite{li2020oscarobjectsemanticsalignedpretraining}, and VinVL
\cite{zhang2021vinvlrevisitingvisualrepresentations} have extensively explored
how cross-modal attention can construct deep, fine-grained semantic alignments
prior to or during the retrieval phase, typically at the cost of an object
detector in the visual pipeline; ViLT
\cite{kim2021viltvisionandlanguagetransformerconvolution} shows that much of
this alignment quality survives when the convolutional backbone is removed
entirely, at a fraction of the inference cost. Separately, the introduction of
hard negative mining in VSE++
\cite{faghri2018vseimprovingvisualsemanticembeddings} substantially improved
discriminative power by concentrating the ranking loss on the most confusing
unmatched samples. Large-scale datasets such as MSCOCO, Flickr30K, and LAION
have further accelerated progress in multimodal retrieval
\cite{lin2015microsoftcococommonobjects,plummer2016flickr30kentitiescollectingregiontophrase,schuhmann2021laion400mopendatasetclipfiltered}.
While highly accurate, these dense retrievers tend to return visually
homogeneous results, creating a critical need for post-hoc diversification.

\section{Preference Reflection Score}
\label{sec:appendix_prs}
Table~\ref{tab:prs_score} reports the full Preference Reflection Score across all datasets and task directions. A score of $1.0$ indicates perfect monotonic alignment between the user-defined $\lambda$ and the resulting diversity metric.

For a controllable retrieval framework to be viable in production systems, it must respond predictably to user-defined intensity constraints. To evaluate this, we utilize the Preference Reflection Score (PRS) introduced by \cite{sogi2025msdppsmultisourcedeterminantalpoint}. The PRS measures the monotonicity of the diversity metric as the user sweeps the preference trade-off parameter ($\lambda$) from 0.0 to 1.0. A score approaching 1.0 indicates perfect alignment: the model smoothly restricts or expands the metadata diversity precisely as requested by the user.

As shown in Table~\ref{tab:prs_score} (Appendix), basic heuristics like MMR 
completely fail at diversity decrease tasks, yielding a PRS of $-1.0000$, 
indicating the model diverged from user intent. The authors of MS-DPP 
\cite{sogi2025msdppsmultisourcedeterminantalpoint} cite the improvement of 
PRS as their primary motivation for introducing complex Tangent Normalization 
(TN) to the Symmetric Positive Definite manifold. 

\mascot{} demonstrates that highly complex manifold normalizations are 
unnecessary to achieve stable controllability. Using the submodular principle 
of diminishing marginal returns (Equation \ref{eq:masmf}), \mascot{} achieves perfect PRS on four of six task-directions: $1.0000$ on both directions for \texttt{PP\_hour}, and $1.0000$ on increase for both \texttt{PP\_geo\_hour} and \texttt{PP\_geo}. On the composite \texttt{PP\_geo\_hour} and single-attribute \texttt{PP\_geo} decrease tasks, \mascot{}'s PRS is lower (0.8005 and 0.6480), reflecting that monotone controllability is harder to maintain when soft-binning must resolve finer or composite metadata. MS-DPP variants retain higher PRS on those two settings, though at substantially lower recall throughout the curve (Figure~\ref{fig:recall_curves}). As the parameter $\lambda$ 
increases, the marginal coverage penalty scales linearly, naturally steering 
the greedy selection algorithm without encountering the gradient explosions 
or matrix instabilities inherent to manifold inversion. \mascot{} achieves comparable controllability to MS-DPP variants on most settings without manifold normalization, and does so at a smaller cost to retrieval quality than manifold inversion incurs (Figure~\ref{fig:recall_curves}).

\begin{table*}[!h]
\begin{center}
\begin{small}
\begin{tabular}{l|cc|cc|cc}
\toprule
& \multicolumn{2}{c|}{\textbf{PP\_geo\_hour}} & \multicolumn{2}{c|}{\textbf{PP\_hour}} & \multicolumn{2}{c}{\textbf{PP\_geo}} \\
\textbf{Method} & \textbf{Inc ($\uparrow$)} & \textbf{Dec ($\uparrow$)} & \textbf{Inc ($\uparrow$)} & \textbf{Dec ($\uparrow$)} & \textbf{Inc ($\uparrow$)} & \textbf{Dec ($\uparrow$)} \\
\midrule
MMR                         & \textbf{1.0000} & -1.0000 & \textbf{1.0000} & -1.0000 & \textbf{1.0000} & -1.0000 \\
k-DPP                       & 0.9991 & 0.8286 & \textbf{1.0000} & -0.5281 & \textbf{1.0000} & \textbf{1.0000} \\
\midrule
MS-DPP                      & 0.9969 & \textbf{0.9768} & 0.9980 & 0.9856 & \textbf{1.0000} & 0.9981 \\
MS-DPP+TN                   & 0.9995 & 0.7717 & \textbf{1.0000} & 0.9276 & \textbf{1.0000} & 0.9953 \\
MS-DPP+TN+TVMS              & 0.9988 & 0.8669 & 0.7434 & 0.9477 & 0.9983 & 0.9937 \\
\midrule
MASCOT (Both Ablated)       & \textbf{1.0000} & 0.8324 & 0.9946 & 0.9870 & 0.2283 & \textbf{1.0000} \\
\textbf{MASCOT (Full)}      & \textbf{1.0000} & 0.8005 & \textbf{1.0000} & \textbf{1.0000} & \textbf{1.0000} & 0.6480 \\
\bottomrule
\end{tabular}
\end{small}
\end{center}
\caption{Preference Reflection Score (PRS) across all datasets. A higher positive score ($\uparrow$) indicates that the model smoothly and accurately adjusts the subset's diversity in alignment with a sweeping user preference parameter ($\lambda \in [0, 1]$). MASCOT attains perfect PRS on four of six task-directions without manifold normalization; on \texttt{PP\_geo\_hour} and \texttt{PP\_geo} decrease it trails the tangent-normalized MS-DPP variants.}
\label{tab:prs_score}
\end{table*}

\section{Datasets}
\label{sec:appendix_datasets}

We evaluate \mascot{} on four datasets that span complementary attribute regimes (shooting time, geographic location, or both). Following \cite{sogi2025msdppsmultisourcedeterminantalpoint}, we use Visual Genome (VG) \cite{krishna2016visualgenomeconnectinglanguage} for tasks involving shooting time, Incidents1M (I1M) \cite{weber2022incidents1mlargescaledatasetimages} for tasks involving shooting location, PixelProse (PP) \cite{singla2024pixelsproselargedataset} for tasks involving either shooting time or location, and additionally include SkyScript \cite{wang2023skyscriptlargesemanticallydiverse} for remote-sensing geographic refinement. Per-dataset split sizes are summarized in Table~\ref{tab:dataset_sizes}.

\begin{table}[!h]
\centering
\begin{small}
\begin{tabular}{lrrr}
\toprule
\textbf{Dataset} & \textbf{Val} & \textbf{Test} & \textbf{Total} \\
\midrule
VG\_hour          & 136     & 547      & 683    \\
I1M\_geo          & 5{,}303 & 21{,}214 & 26{,}517 \\
PP (all variants) & 199     & 797      & 996    \\
SkyScript (all)   & 3{,}130 & 12{,}520 & 15{,}650 \\
\bottomrule
\end{tabular}
\end{small}
\caption{Per-dataset validation and test split sizes used in our experiments. All datasets use a 20\%/80\% val/test split over the retained images.}
\label{tab:dataset_sizes}
\end{table}

\paragraph{PixelProse (PP)} PixelProse \cite{singla2024pixelsproselargedataset} is a large-scale dataset of over 16M images with dense synthetic captions generated by a vision-language model, sourced from CommonPool, CC12M, and RedCaps. Following \cite{sogi2025msdppsmultisourcedeterminantalpoint}, we retain only images whose EXIF metadata contains both valid \texttt{GPSLatitude}/\texttt{GPSLongitude} and \texttt{DateTimeOriginal} fields. On the current HuggingFace release of PixelProse, this metadata filter yields 1{,}799 candidate images, of which 996 are successfully downloaded from their source URLs; the remainder are lost to URL attrition across CommonPool, CC12M, and RedCaps hosts. The 996 retained images are substantially fewer than the 25{,}151 reported in \cite{sogi2025msdppsmultisourcedeterminantalpoint}, a gap we attribute to updates in the PixelProse HuggingFace release subsequent to their work and to continued decay of source image URLs. Because the same 996 images are used for both \mascot{} and all baselines, relative comparisons remain valid even though absolute image counts differ from the original MS-DPP setup. Latitude/longitude pairs are converted to unit 3D vectors on the sphere for the Vendi-based diversity metric, and hour/minute values are mapped to cyclical time embeddings; soft-binning itself operates on the raw degree coordinates (Section~\ref{sec:framework}). For text queries, we randomly sample 1{,}000 dense captions; each caption serves as the single ground-truth match for exactly one image. PP supports all three composite-attribute tasks: \texttt{PP\_geo}, \texttt{PP\_hour}, and \texttt{PP\_geo\_hour}.

\paragraph{Visual Genome (VG)} Visual Genome \cite{krishna2016visualgenomeconnectinglanguage} contains over 100K natural images densely annotated with objects, attributes, and relationships. We use the lightweight \texttt{image\_data.json} and \texttt{objects.json} metadata files rather than the full image archive. Because the vast majority of VG images lack EXIF timestamps, we scan all candidate images using HTTP range requests (fetching only the first 64KB of each JPEG, sufficient for EXIF headers) and retain only those exposing a valid \texttt{DateTimeOriginal} tag with parseable hour/minute fields. This yields 683 images with reliable shooting-time metadata, closely matching the 649 reported in \cite{sogi2025msdppsmultisourcedeterminantalpoint}; small variations in download success rates for VG source URLs account for the minor discrepancy. For text queries, we use the fifteen dominant object names: \textit{person, building, table, vehicle, animal, tree, sky, road, chair, light, car, desk, bird, apple, dog}, with positives derived from VG's object annotations. VG supports the \texttt{VG\_hour} task.

\paragraph{Incidents1M (I1M)} Incidents1M \cite{weber2022incidents1mlargescaledatasetimages} is a multi-label dataset of 977K images depicting natural disasters and incidents, annotated with 43 incident categories and 49 place categories. We use the multi-label val subset and retain only entries with at least one positive incident label. Since I1M does not provide explicit GPS coordinates, we follow the methodology of the original Incidents1M paper and approximate shooting locations via the image URL: each unique domain is resolved to an IP address through DNS lookup, then geolocated using public IP-geolocation APIs (\texttt{geolocation-db.com} and \texttt{ip-api.com}). Domains failing to resolve to valid (non-zero) coordinates are discarded. To prevent identical coordinates from collapsing into a single point, we add small uniform noise (scaled by the duplicate count) to images sharing the same rounded location. Latitude/longitude pairs are then converted to unit 3D vectors for the diversity metric, with soft-binning again operating on degrees. After filtering for images that download successfully and possess valid GPS, we obtain 26{,}517 images, slightly more than the 22{,}547 reported in \cite{sogi2025msdppsmultisourcedeterminantalpoint}; the difference reflects drift in domain-to-IP mappings and updates to the underlying geolocation databases between runs. For text queries, we use the disaster types and shooting-location types originally annotated in I1M, filtered to those with at least two positive examples in each split. I1M supports the \texttt{I1M\_geo} task.

\paragraph{SkyScript} SkyScript \cite{wang2023skyscriptlargesemanticallydiverse} is a remote-sensing vision-language dataset of 2.6M image-caption pairs covering 29K distinct OpenStreetMap semantic tags, with images sourced from Google Earth Engine. We use the SkyScript\_test\_30K split (CLIP-filtered test set) and download the corresponding image and metadata pickles directly from the SkyScript S3 bucket via remote zip readers. Each image's metadata pickle provides a bounding box, from which we compute the center latitude/longitude as the shooting location, normalized to a unit 3D vector for the diversity metric. We retain only images with valid GPS, successfully downloaded JPEG content, and unique \texttt{title\_multi\_objects} captions (which yield substantially higher caption diversity than the focus-object \texttt{title} field), yielding 15{,}650 images. For text queries, each retained caption serves as the single ground-truth match for its corresponding image. SkyScript supports the \texttt{SkyScript\_geo} task.

\paragraph{Splits and queries} For each dataset, we use 20\% of images as the validation set and the remaining 80\% as the test set. Hyperparameters (diversification intensity $\lambda$, soft-binning bandwidths $\sigma_{\text{geo}}$ and $\sigma_{\text{time}}$, and geographic grid resolution) are tuned on the validation set via grid search and held fixed at test time. For VG and I1M, where each query can have multiple positive images, queries with fewer than two positives in either split are dropped, and only queries present in both val and test splits are retained for evaluation. For PP and SkyScript, each caption defines a single-ground-truth retrieval problem, with each query caption corresponding to exactly one image.

\section{Sensitivity Analysis}
\label{sec:appendix_sensitivity}
This appendix provides the full sensitivity sweeps. We analyze three hyperparameters: diversification intensity $\lambda$, soft-binning bandwidths $\sigma_{\text{geo}}$ and $\sigma_{\text{time}}$, and geographic grid resolution.

Because the parameters interact, each sweep must be anchored somewhere. The $\lambda$ sweeps below are run at each task-direction's own validation-selected bandwidths, listed in Table~\ref{tab:hyperparams}, so that the reported curve passes through the operating point used in Tables~\ref{tab:decrease} and~\ref{tab:increase}. The bandwidth and grid-resolution sweeps hold $\lambda$ at each method's selected value.

The two parameters are not independent. Sweeping $\lambda$ at each task's selected bandwidth shows that the transition from relevance-preserving to relevance-destroying behavior is a $\sigma$-dependent boundary rather than a fixed range of $\lambda$. On \texttt{PP\_geo} decrease ($\sigma_{\text{geo}} = 10$), R@10 holds at 0.8105 for $\lambda = 0.3$ and falls to zero by $\lambda = 0.5$. On \texttt{PP\_hour} decrease ($\sigma_{\text{time}} = 0.5$), it holds at 0.9059 for $\lambda = 0.4$ and falls to 0.5307 by $\lambda = 0.5$. On the composite \texttt{PP\_geo\_hour} decrease task ($\sigma_{\text{geo}} = 15$, $\sigma_{\text{time}} = 3.0$) the boundary is tighter still: R@10 = 0.9410 at $\lambda = 0.1$ and 0.0477 at $\lambda = 0.2$. Wider bandwidths shift the boundary toward smaller $\lambda$, because diffuse soft assignments cause the coverage penalty to fire on nearly every candidate before any bin saturates.

We state plainly that this is a limitation, and it is not confined to $\lambda$. Taking one grid step past the selected value along each axis in turn, R@10 on \texttt{PP\_geo} decrease falls from 0.8105 to 0.0753 ($\lambda$), 0.0013 ($\sigma_{\text{geo}}$) and 0.0025 (grid resolution); on \texttt{PP\_hour} decrease from 0.9059 to 0.5307 ($\lambda$) and 0.0351 ($\sigma_{\text{time}}$); on \texttt{PP\_geo\_hour} decrease from 0.9410 to 0.0477, 0.4053 and 0.5834 respectively. The exception is $\sigma_{\text{time}}$ on the composite task, where R@10 stays above 0.93 across the entire range, since the geographic sub-space dominates the coverage penalty there. Validation grid search alone does not indicate how close a chosen operating point is to these boundaries. Practitioners should sweep each discretization parameter at their chosen setting and select values strictly interior to the stable region, accepting weaker concentration in exchange for margin. One property makes this practical: collapse is abrupt and near-total rather than gradual, so a single probe one step past the intended value reliably detects proximity.

The mechanism is the one that produces the method's advantage. The coverage penalty decays only as bins saturate; when bandwidth and grid resolution are such that bins rarely saturate, the penalty instead fires on every selection and suppresses relevant candidates indiscriminately.
The same effect appears along the grid-resolution axis, where R@10 on \texttt{PP\_geo} decrease falls from 0.9686 at $g = 5$ to 0.8105 at the selected $g = 20$ and to 0.0025 at $g = 25$. Increase tasks are far more forgiving: on \texttt{PP\_geo} increase, R@10 remains above 0.86 through $\lambda = 0.9$, since rewarding coverage never starves the relevance term in the same way.

\subsection{Per-Task Hyperparameters}
\label{sec:appendix_hyperparams}

Table~\ref{tab:hyperparams} lists the validation-selected hyperparameters for every (task, direction, method) triple reported in the main tables. Retrieval and diversity values are reproduced from Tables~\ref{tab:decrease}, \ref{tab:increase}, \ref{tab:vg_hour_results} and~\ref{tab:i1m_geo_results}; a fresh run at each listed configuration reproduces them to within $10^{-3}$. Recall that $\sigma_{\text{geo}}$ is in degrees and $\sigma_{\text{time}}$ in hours (Section~\ref{sec:framework}), and that geographic tasks use a $20 \times 20$ grid. PixelProse and Visual Genome report R@10; Incidents1M reports MAP.

\begin{table*}[!h]
\centering
\begin{small}
\begin{tabular}{llccccc}
\toprule
\textbf{Task} & \textbf{Dir} & \textbf{Method} & $\lambda$ & $\sigma_{\text{geo}}$ & $\sigma_{\text{time}}$ & \textbf{R@10 / MAP} \\
\midrule
\multirow{6}{*}{\texttt{PP\_geo}}
 & dec & \mascot{}          & 0.3 & 10.0 & 0.5 & 0.8105 \\
 & dec & Uniform Binning    & 0.5 & 10.0 & 0.5 & 0.8695 \\
 & dec & w/o Normalization  & 0.5 &  5.0 & 0.5 & 0.5947 \\
 & inc & \mascot{}          & 0.8 & 15.0 & 0.5 & 0.9109 \\
 & inc & Uniform Binning    & 0.5 & 10.0 & 0.5 & 0.8821 \\
 & inc & w/o Normalization  & 0.9 &  1.0 & 0.5 & 0.9398 \\
\midrule
\multirow{6}{*}{\texttt{PP\_hour}}
 & dec & \mascot{}          & 0.4 & 1.0 & 0.5 & 0.9059 \\
 & dec & Uniform Binning    & 0.7 & 1.0 & 0.5 & 0.8381 \\
 & dec & w/o Normalization  & 0.5 & 1.0 & 0.5 & 0.2974 \\
 & inc & \mascot{}          & 0.9 & 1.0 & 1.5 & 0.8921 \\
 & inc & Uniform Binning    & 0.7 & 1.0 & 0.5 & 0.8946 \\
 & inc & w/o Normalization  & 0.5 & 1.0 & 0.5 & 0.9072 \\
\midrule
\multirow{6}{*}{\texttt{PP\_geo\_hour}}
 & dec & \mascot{}          & 0.1 & 15.0 & 3.0 & 0.9410 \\
 & dec & Uniform Binning    & 0.7 & 10.0 & 0.5 & 0.8118 \\
 & dec & w/o Normalization  & 0.5 &  1.0 & 0.5 & 0.2873 \\
 & inc & \mascot{}          & 0.8 & 15.0 & 1.5 & 0.8356 \\
 & inc & Uniform Binning    & 0.5 &  1.0 & 0.5 & 0.9072 \\
 & inc & w/o Normalization  & 0.5 &  1.0 & 0.5 & 0.8871 \\
\midrule
\multirow{6}{*}{\texttt{VG\_hour}}
 & dec & \mascot{}          & 0.4 & --- & 0.5 & 0.9333 \\
 & dec & Uniform Binning    & 0.9 & --- & 0.5 & 0.8667 \\
 & dec & w/o Normalization  & 0.5 & --- & 0.5 & 0.8667 \\
 & inc & \mascot{}          & 0.7 & --- & 1.5 & 0.8667 \\
 & inc & Uniform Binning    & 0.7 & --- & 0.5 & 0.9333 \\
 & inc & w/o Normalization  & 0.3 & --- & 0.5 & 0.9333 \\
\midrule
\multirow{6}{*}{\texttt{I1M\_geo}}
 & dec & \mascot{}          & 0.7 &  5.0 & --- & 0.6949 \\
 & dec & Uniform Binning    & 0.7 & 10.0 & --- & 0.6781 \\
 & dec & w/o Normalization  & 0.1 & 10.0 & --- & 0.6521 \\
 & inc & \mascot{}          & 0.7 & 10.0 & --- & 0.7160 \\
 & inc & Uniform Binning    & 0.1 & 10.0 & --- & 0.7333 \\
 & inc & w/o Normalization  & 0.1 & 10.0 & --- & 0.7178 \\
\bottomrule
\end{tabular}
\end{small}
\caption{Validation-selected hyperparameters per task, direction and method. ``---'' marks a parameter not applicable to the task's attribute type. \texttt{I1M\_geo} reports MAP; all other tasks report R@10.}
\label{tab:hyperparams}
\end{table*}

\subsection{Diversification Intensity ($\lambda$)}
 
\begin{table}[!h]
\begin{center}
\begin{small}
\setlength{\tabcolsep}{3.5pt}
\begin{tabular}{lcccccc}
\toprule
& \multicolumn{2}{c}{\textbf{PP\_geo}} & \multicolumn{2}{c}{\textbf{PP\_hour}} & \multicolumn{2}{c}{\textbf{PP\_geo\_hour}} \\
\cmidrule(lr){2-3}\cmidrule(lr){4-5}\cmidrule(lr){6-7}
$\lambda$ & Dec & Inc & Dec & Inc & Dec & Inc \\
\midrule
0.0 & 0.9737 & 0.9737 & 0.9737 & 0.9737 & 0.9737 & 0.9737 \\
0.1 & 0.9737 & 0.9737 & 0.9724 & 0.9749 & \textbf{0.9410} & 0.9737 \\
0.2 & 0.9586 & 0.9686 & 0.9737 & 0.9737 & 0.0477 & 0.9674 \\
0.3 & \textbf{0.8105} & 0.9649 & 0.9598 & 0.9686 & 0.0088 & 0.9573 \\
0.4 & 0.0753 & 0.9561 & \textbf{0.9059} & 0.9674 & 0.0013 & 0.9410 \\
0.5 & 0.0000 & 0.9511 & 0.5307 & 0.9649 & 0.0013 & 0.9222 \\
0.7 & 0.0000 & 0.9348 & 0.0063 & 0.9523 & 0.0000 & 0.8770 \\
0.8 & 0.0000 & \textbf{0.9109} & 0.0013 & 0.9373 & 0.0000 & \textbf{0.8356} \\
0.9 & 0.0000 & 0.8607 & 0.0000 & \textbf{0.8921} & 0.0000 & 0.7955 \\
1.0 & 0.0000 & 0.3752 & 0.0000 & 0.1330 & 0.0000 & 0.6236 \\
\bottomrule
\end{tabular}
\end{small}
\end{center}
\caption{\mascot{} R@10 as $\lambda$ sweeps, with $\sigma$ held at each task-direction's validation-selected values (Table~\ref{tab:hyperparams}). Bold marks the operating point of Tables~\ref{tab:decrease} and~\ref{tab:increase}.}
\label{tab:sensitivity_theta}
\end{table}

At $\lambda = 0$ the objective reduces to the unconstrained BLIP-2 retriever (R@10 = 0.9737) on every task. Increase and decrease behave very differently thereafter. Increase tasks degrade gradually, retaining above 0.86 R@10 through $\lambda = 0.9$ on \texttt{PP\_geo} and above 0.79 on \texttt{PP\_geo\_hour}, collapsing only when relevance is entirely suppressed at $\lambda = 1.0$. Decrease tasks instead exhibit an abrupt boundary whose location depends on bandwidth: $\lambda \approx 0.4$ on \texttt{PP\_geo}, $\lambda \approx 0.5$ on \texttt{PP\_hour}, and $\lambda \approx 0.15$ on the composite task, where the wider bandwidths ($\sigma_{\text{geo}} = 15$, $\sigma_{\text{time}} = 3.0$) make bins slow to saturate. In every decrease case the reported operating point is the last value before collapse.


\subsection{Geographic Soft-Binning Bandwidth ($\sigma_{\text{geo}}$)}
 
\begin{table}[!h]
\begin{center}
\begin{small}
\begin{tabular}{lcccc}
\toprule
& \multicolumn{2}{c}{\textbf{PP\_geo}} & \multicolumn{2}{c}{\textbf{PP\_geo\_hour}} \\
\cmidrule(lr){2-3}\cmidrule(lr){4-5}
$\sigma_{\text{geo}}$ & Dec & Inc & Dec & Inc \\
\midrule
0.5  & 0.9737 & 0.9737 & 0.9699 & 0.9373 \\
1.0  & 0.9737 & 0.9737 & 0.9699 & 0.9310 \\
2.0  & 0.9737 & 0.9624 & 0.9699 & 0.8808 \\
5.0  & 0.9649 & 0.8319 & 0.9711 & 0.7453 \\
10.0 & \textbf{0.8105} & 0.9260 & 0.9686 & 0.8557 \\
15.0 & 0.0013 & \textbf{0.9109} & \textbf{0.9410} & \textbf{0.8356} \\
20.0 & 0.0000 & 0.8858 & 0.4053 & 0.8269 \\
30.0 & 0.0151 & 0.8419 & 0.0728 & 0.7779 \\
\bottomrule
\end{tabular}
\end{small}
\end{center}
\caption{\mascot{} R@10 as $\sigma_{\text{geo}}$ sweeps, other parameters at each task-direction's validation-selected values. Grid size fixed at 20.}
\label{tab:sensitivity_sigma_geo}
\end{table}
 
Geographic bandwidth behaves differently by direction. On decrease tasks, small bandwidths keep each image's assignment localized and the coverage penalty concentrated: R@10 on \texttt{PP\_geo} holds at 0.9737 for $\sigma_{\text{geo}} \leq 2.0$ and 0.9649 at 5.0, then falls to 0.8105 at the selected value of 10.0 and to 0.0013 at 15.0. \texttt{PP\_geo\_hour} tolerates wider bandwidths, above 0.94 through $\sigma_{\text{geo}} = 15.0$, because its coverage term is split across geographic and temporal sub-spaces, so diffuse geographic assignments are partially offset by sharper temporal ones. Both tasks collapse once bins overlap enough that none saturates within the $K = 20$ budget. Increase tasks are markedly more robust, retaining R@10 above 0.74 across the entire range on both tasks, since rewarding coverage never starves the relevance term.

\subsection{Temporal Soft-Binning Bandwidth ($\sigma_{\text{time}}$)}
 
\begin{table}[!h]
\begin{center}
\begin{small}
\begin{tabular}{lcccc}
\toprule
& \multicolumn{2}{c}{\textbf{PP\_hour}} & \multicolumn{2}{c}{\textbf{PP\_geo\_hour}} \\
\cmidrule(lr){2-3}\cmidrule(lr){4-5}
$\sigma_{\text{time}}$ & Dec & Inc & Dec & Inc \\
\midrule
0.25 & 0.9172 & 0.3588 & 0.9561 & 0.7077 \\
0.5  & \textbf{0.9059} & 0.7955 & 0.9511 & 0.8482 \\
1.0  & 0.0351 & 0.9021 & 0.9398 & 0.8507 \\
1.5  & 0.0527 & \textbf{0.8921} & 0.9348 & \textbf{0.8356} \\
2.0  & 0.0602 & 0.9072 & 0.9385 & 0.8331 \\
3.0  & 0.0602 & 0.9536 & \textbf{0.9410} & 0.8394 \\
5.0  & 0.7716 & 0.9661 & 0.9398 & 0.8683 \\
\bottomrule
\end{tabular}
\end{small}
\end{center}
\caption{\mascot{} R@10 as $\sigma_{\text{time}}$ sweeps, other parameters at each task-direction's validation-selected values. Number of temporal bins fixed at 24.}
\label{tab:sensitivity_sigma_time}
\end{table}

Temporal bandwidth shows the sharpest interaction with task structure. On \texttt{PP\_hour} decrease, where time is the only attribute, R@10 falls from 0.9059 at $\sigma_{\text{time}} = 0.5$ to 0.0351 at 1.0 (a 96\% drop in one step) because a bandwidth spanning several hourly bins prevents any single bin from saturating. The partial recovery at $\sigma_{\text{time}} = 5.0$ (0.7716) occurs for the opposite reason: assignments become so diffuse that every bin appears near-uniformly covered from the outset and the penalty is nearly constant across candidates, leaving relevance to dominate. On \texttt{PP\_geo\_hour} decrease the same sweep is essentially flat (0.9348--0.9561 throughout), since the 400-cell geographic sub-space contributes the bulk of the coverage signal and the 24 temporal bins cannot destabilize it. This asymmetry is the clearest evidence that the stability boundary is a property of the combined Information Unit space rather than of any single bandwidth.

\subsection{Geographic Grid Resolution}
 
\begin{table}[!h]
\begin{center}
\begin{small}
\begin{tabular}{lcccc}
\toprule
& \multicolumn{2}{c}{\textbf{PP\_geo}} & \multicolumn{2}{c}{\textbf{PP\_geo\_hour}} \\
\cmidrule(lr){2-3}\cmidrule(lr){4-5}
Grid size $g$ & Dec & Inc & Dec & Inc \\
\midrule
5  & 0.9686 & 0.9649 & 0.9661 & 0.8770 \\
10 & 0.9661 & 0.9247 & 0.9724 & 0.8193 \\
15 & 0.9448 & 0.9210 & 0.9661 & 0.8482 \\
20 & \textbf{0.8105} & \textbf{0.9109} & \textbf{0.9410} & \textbf{0.8356} \\
25 & 0.0025 & 0.8996 & 0.5834 & 0.8218 \\
30 & 0.0000 & 0.8620 & 0.0715 & 0.8118 \\
40 & 0.0000 & 0.7992 & 0.0063 & 0.7905 \\
50 & 0.0000 & 0.7541 & 0.0013 & 0.7779 \\
\bottomrule
\end{tabular}
\end{small}
\end{center}
\caption{\mascot{} R@10 as geographic grid resolution sweeps, other parameters at each task-direction's validation-selected values.}
\label{tab:sensitivity_grid_size}
\end{table}

Grid resolution controls how many geographic cells the same coordinates are spread across; $g$ denotes a $g \times g$ partition, so $g = 5$ yields 25 continental-scale cells and $g = 50$ yields 2500 narrow ones. On decrease tasks, coarse grids are strictly safer: each large cell contains many candidates, the relevant bin saturates after one or two selections, and subsequent selections pay almost no penalty. \texttt{PP\_geo} holds R@10 above 0.94 through $g = 15$ and \texttt{PP\_geo\_hour} above 0.96 through $g = 15$, with both collapsing beyond the selected $g = 20$ to 0.0025 and 0.5834 at $g = 25$, and to near zero by $g = 30$. At fine resolutions each image occupies a nearly unique cell, so bins never saturate and the penalty fires on every selection. Increase tasks again degrade only gradually (0.9649 to 0.7541 on \texttt{PP\_geo} across the full range), because the query-driven weights concentrate the retrieval budget on populated cells regardless of how many cells exist.

\subsection{Fixed-Recall Comparison}
\label{sec:appendix_fixed_recall}
Harmonic means conflate two quantities that practitioners often wish to constrain separately. Table~\ref{tab:fixed_recall} therefore reports the best diversity metric each method achieves subject to a floor on R@10, on the composite \texttt{PP\_geo\_hour} decrease task.

\begin{table}[!h]
\centering
\begin{small}
\begin{tabular}{lccc}
\toprule
\textbf{Method} & \textbf{R@10 $\geq$ 0.95} & \textbf{R@10 $\geq$ 0.90} & \textbf{R@10 $\geq$ 0.85} \\
\midrule
BLIP-2 (Base)  & 0.1656 & 0.1656 & 0.1656 \\
MS-DPP         & ---    & ---    & ---    \\
MS-DPP+TN+TVMS & \textbf{0.1902} & \textbf{0.2066} & 0.2189 \\
Prob-Coverage  & 0.1656 & 0.1656 & \textbf{0.2860} \\
\mascot{}      & 0.1784 & 0.1881 & 0.1881 \\
\bottomrule
\end{tabular}
\end{small}
\caption{Maximum DM attainable subject to an R@10 floor, \texttt{PP\_geo\_hour} decrease. ``---'' indicates no operating point meets the floor; MS-DPP's best R@10 on this task is 0.4931.}
\label{tab:fixed_recall}
\end{table}

Two observations follow. MS-DPP cannot meet any of these floors at all, which is the practical form of the failure mode this paper identifies. Among methods that can, tangent-normalized MS-DPP concentrates slightly harder than \mascot{} at the two tighter floors, and Prob-Coverage leads at the loosest. \mascot{} is thus not the strongest compressor at matched recall; its advantage is that it reaches high-recall operating points at all under composite constraints, which vanilla MS-DPP does not.

\section{Early-Rank Semantic Integrity on VG and I1M}
\label{sec:appendix_recall_curves}
 
\begin{figure*}[t]
    \centering
    \begin{subfigure}{0.45\textwidth}
        \includegraphics[width=\linewidth]{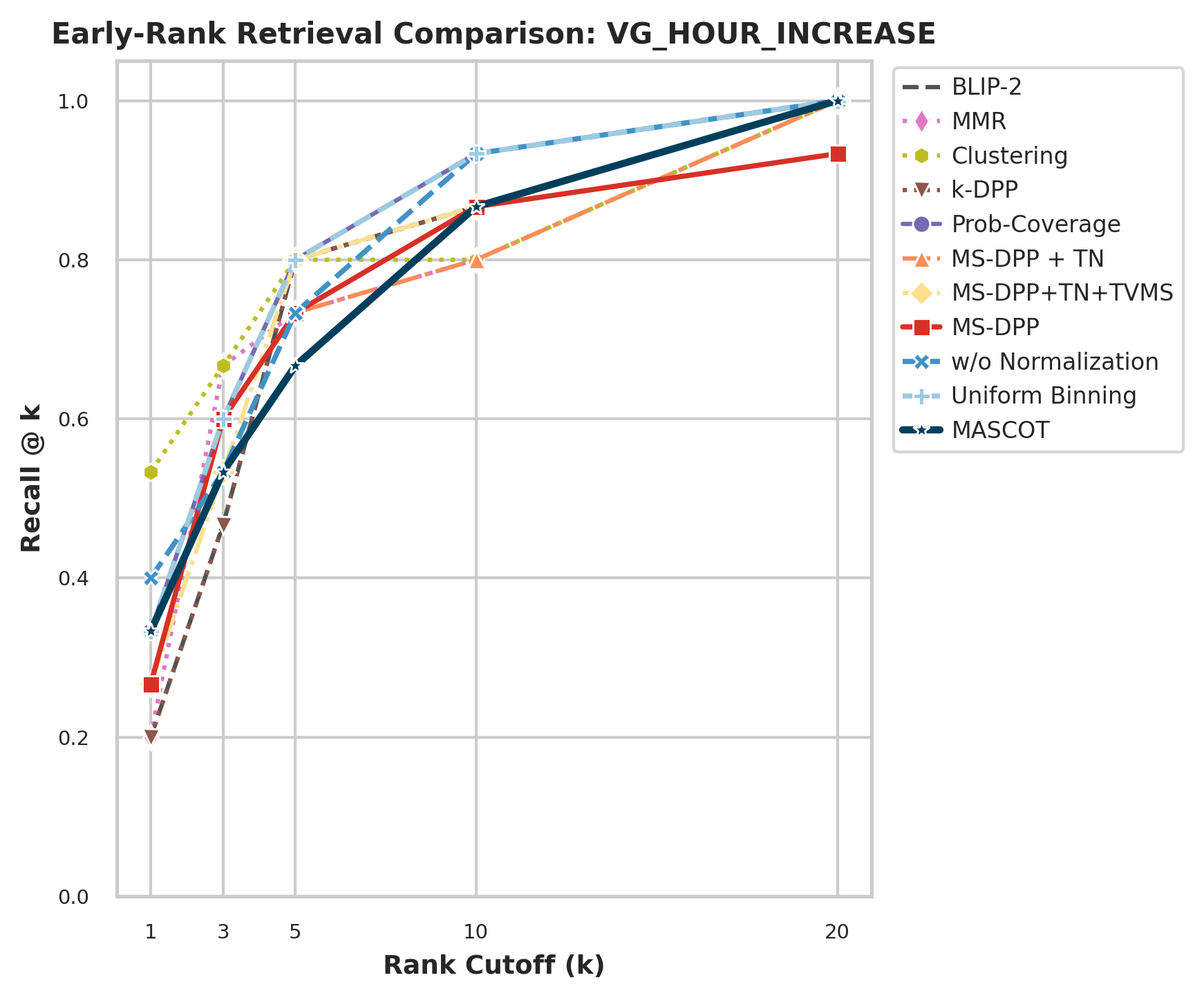}
        \caption{VG Hour (Increase)}
    \end{subfigure}\hfill
    \begin{subfigure}{0.45\textwidth}
        \includegraphics[width=\linewidth]{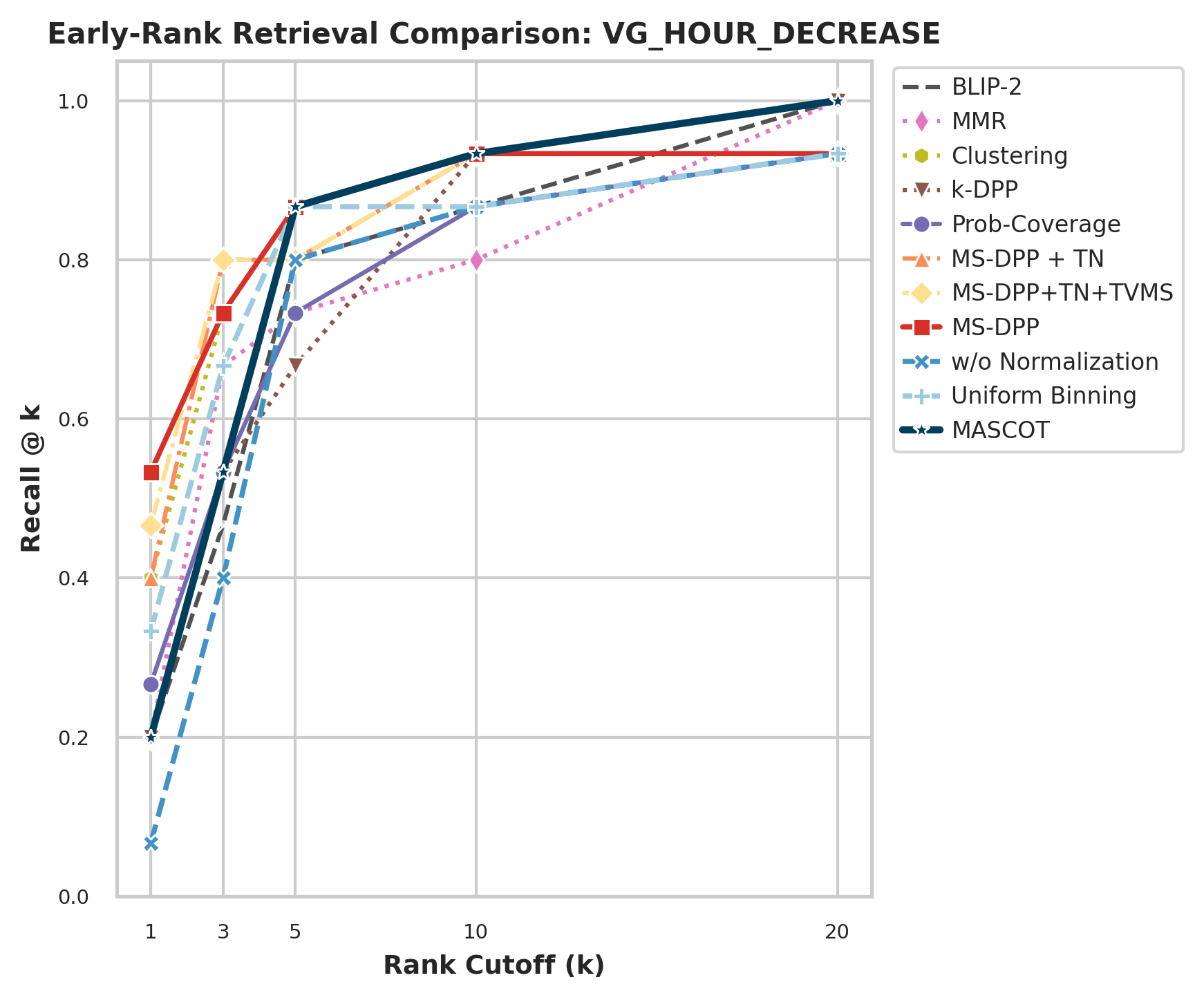}
        \caption{VG Hour (Decrease)}
    \end{subfigure}
 
    \vspace{4mm}
 
    \begin{subfigure}{0.45\textwidth}
        \includegraphics[width=\linewidth]{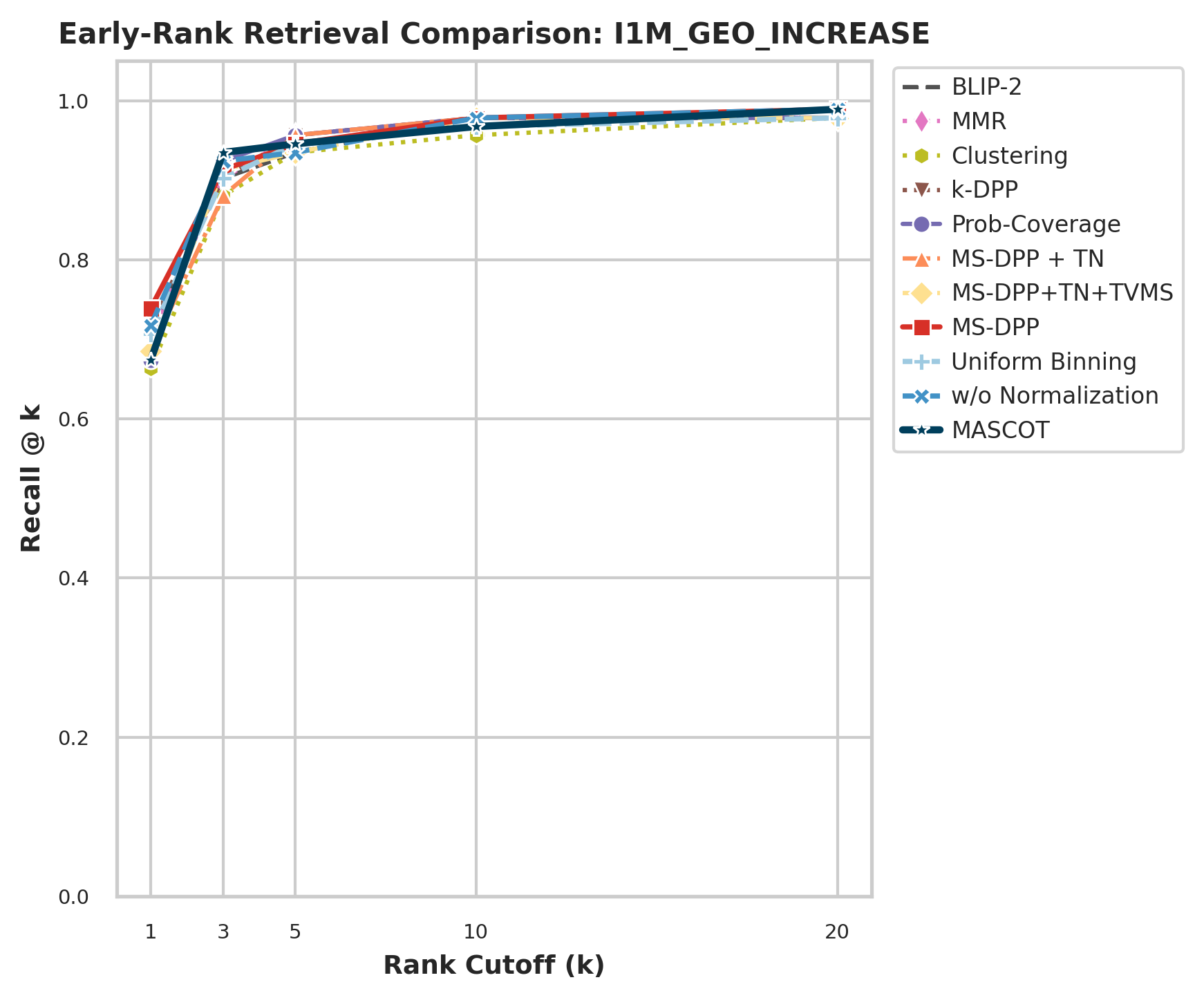}
        \caption{I1M Geo (Increase)}
    \end{subfigure}\hfill
    \begin{subfigure}{0.45\textwidth}
        \includegraphics[width=\linewidth]{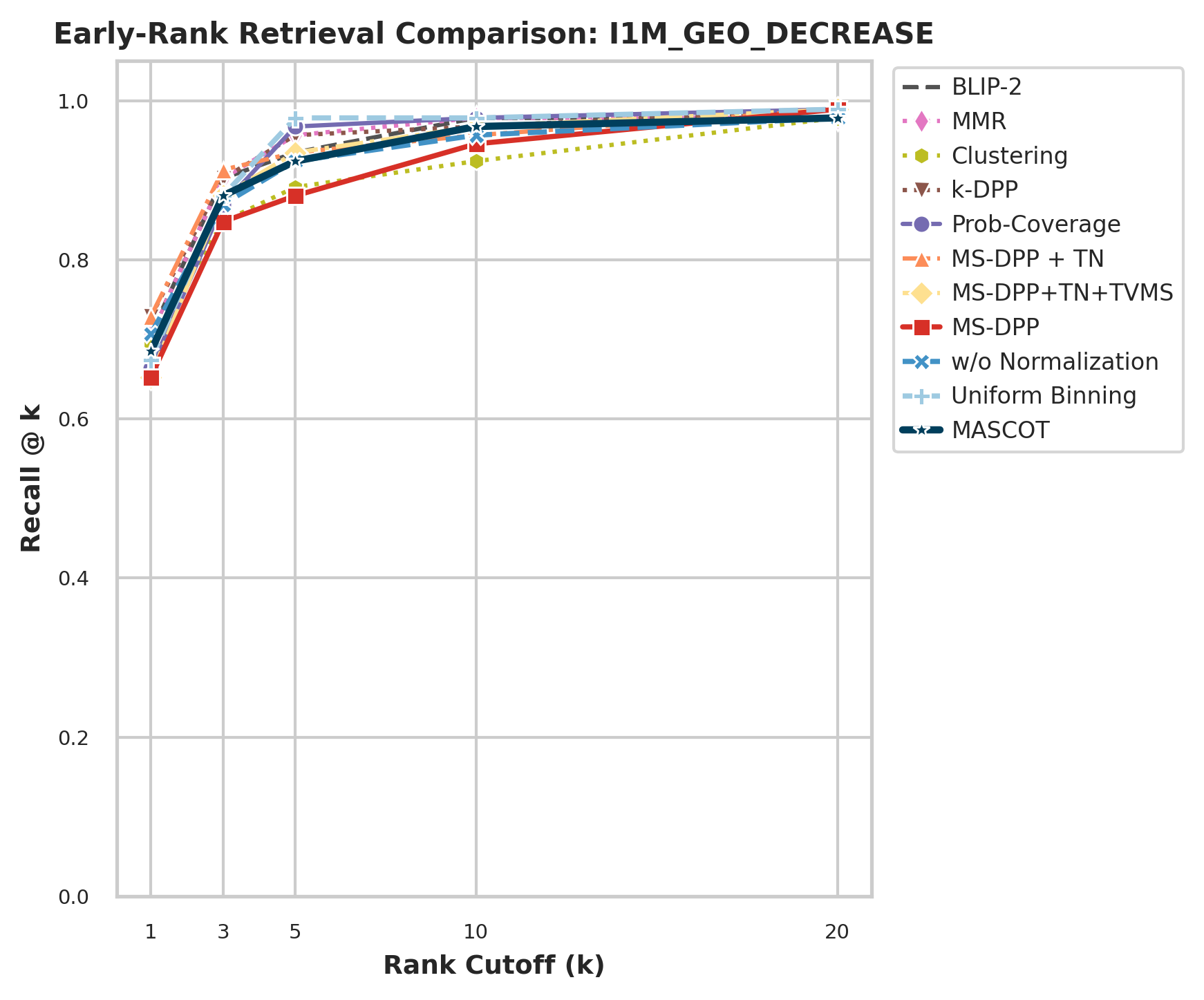}
        \caption{I1M Geo (Decrease)}
    \end{subfigure}
 
    \caption{Recall@K curves ($K \in \{1, 3, 5, 10, 20\}$) on the \texttt{VG\_hour} and \texttt{I1M\_geo} benchmarks. \textbf{Left column (Increase):} All methods track closely, consistent with the PixelProse findings; spatial repulsion and coverage-based models alike preserve semantic recall during diversity expansion. \textbf{Right column (Decrease):} The severity of the manifold vulnerability depends on dataset characteristics, and on \texttt{VG\_hour} it does not appear at all. With 15 queries over 683 images, R@1 takes discrete values in steps of $1/15 \approx 0.067$ and is correspondingly noisy: the unconstrained BLIP-2 baseline itself reaches only 0.2000 at rank~1, which \mascot{} matches exactly and which every MS-DPP variant exceeds (0.4000--0.5333). The candidate pool is too small for manifold repulsion to overreach, so no early-rank collapse occurs and this setting is uninformative about the mechanism we study. On \texttt{I1M\_geo} (26K images), MS-DPP and MS-DPP+TN+TVMS both fall to R@1 = 0.6522 against the BLIP-2 baseline's 0.7065, while \mascot{} (0.6848) and k-DPP (0.7283) stay closer; the margin is small, consistent with I1M's IP-derived coordinates limiting how far any metadata-driven method can separate selections. Together these results bound the early-rank behavior demonstrated on PixelProse rather than extending it.}
    \label{fig:recall_curves_vg_i1m}
\end{figure*}

To complement the early-rank analysis presented in Section~\ref{sec:results} on PixelProse, we report Recall@K curves ($K \in \{1, 3, 5, 10, 20\}$) for the \texttt{VG\_hour} and \texttt{I1M\_geo} benchmarks in Figure~\ref{fig:recall_curves_vg_i1m}.
 
\section{Extended Results on VG, I1M, and SkyScript}
\label{sec:appendix_extended_results}

The advantage of coverage over manifold repulsion is not uniform, and characterizing its boundary is important for practitioners. On the two additional benchmarks, MS-DPP achieves a higher overall harmonic mean than \mascot{} on decrease tasks: on \texttt{VG\_hour} (0.6265 vs.\ 0.5441) and on \texttt{I1M\_geo} (0.5587 vs.\ 0.4973, computed on MAP rather than R@10 since I1M queries have many positives). Two dataset properties explain this. \texttt{VG\_hour} is small, 15 query types over 683 images, so the candidate pool is too limited for manifold repulsion to collapse in the way it does at PixelProse scale; the pathology we exploit simply does not manifest. \texttt{I1M\_geo} derives its coordinates from server-IP geolocation rather than capture GPS, so many images from one news domain share near-identical locations, forming dense clusters that fall into the same soft-binning bin regardless of bandwidth; coverage cannot separate what the metadata does not distinguish. On \texttt{SkyScript\_geo}, \mascot{} attains the highest overall score among diversification methods in both directions, though on increase it is marginally below the unconstrained baseline and at a near-zero absolute scale where all methods are effectively tied. The pattern across benchmarks is consistent: coverage-based selection helps most when metadata is capture-accurate and the candidate pool is large enough for manifold repulsion to overreach, the regime that motivates this work, and offers no advantage when either condition fails.

Tables~\ref{tab:vg_hour_results}, \ref{tab:i1m_geo_results}, and \ref{tab:skyscript_geo_results} report results on the Visual Genome (\texttt{VG\_hour}), Incidents1M (\texttt{I1M\_geo}), and SkyScript (\texttt{SkyScript\_geo}) benchmarks. Taken together, these datasets reveal the boundary conditions of \mascot{}'s coverage-based approach and contextualize the PixelProse findings within a broader range of dataset scales and attribute types.
 
\paragraph{VG\_hour.} The Visual Genome dataset presents a uniquely constrained evaluation setting: with only 15 queries (one per dominant object category) and 683 images, the R@10 metric takes discrete values in steps of $1/15 \approx 0.067$, which sharply limits its discriminative power. Despite this, interpretable patterns emerge. For diversity increase tasks, our coverage-based ablations (Uniform Binning, Prob-Coverage) achieve the highest harmonic means (0.9276 and 0.9270), outperforming MS-DPP variants. \mascot{} remains competitive (0.8772) but does not lead, likely due to the small corpus size (683 images). For diversity decrease tasks, Uniform Binning achieves the best overall score (0.6300), with MS-DPP (0.6265) close behind. This is attributable to the small corpus size: with only 683 candidate images, the mechanism we identify does not manifest, so this benchmark does not discriminate between the two approaches. \mascot{} scores 0.5441, reflecting the inherent difficulty of enforcing strict temporal concentration when the candidate pool is limited.

\begin{table*}[!h]
\begin{center}
\begin{small}
\begin{tabular}{lcc}
\toprule
& \textbf{VG\_hour (Increase)} & \textbf{VG\_hour (Decrease)} \\
\textbf{Method} & \textbf{HM$\uparrow$(R@10$\uparrow$, DM$\uparrow$)} & \textbf{HM$\uparrow$(R@10$\uparrow$, DM$\uparrow$)} \\
\midrule
BLIP-2 (Base)   & 0.8697 (0.8667, 0.8728) & 0.4222 (0.8667, 0.2791) \\
\midrule
Clustering      & 0.8418 (0.8000, 0.8882) & 0.5152 (0.9333, 0.3558) \\
MMR             & 0.8386 (0.8000, 0.8811) & 0.3994 (0.8000, 0.2661) \\
k-DPP           & 0.8809 (0.8667, 0.8955) & 0.4745 (0.9333, 0.3181) \\
\midrule
MS-DPP          & 0.8910 (0.8667, 0.9167) & \underline{0.6265} (0.9333, 0.4714) \\
MS-DPP+TN       & 0.8499 (0.8000, 0.9064) & 0.5795 (0.9333, 0.4202) \\
MS-DPP+TN+TVMS  & 0.8783 (0.8667, 0.8902) & 0.5846 (0.9333, 0.4256) \\
\midrule
\textbf{Ours (Prob-Coverage)}     & \underline{0.9270} (0.9333, 0.9208) & 0.6065 (0.8667, 0.4664) \\
\textbf{Ours (w/o Normalization)} & 0.9207 (0.9333, 0.9085) & 0.6075 (0.8667, 0.4677) \\
\textbf{Ours (Uniform Binning)}   & \textbf{0.9276} (0.9333, 0.9219) & \textbf{0.6300} (0.8667, 0.4948) \\
\textbf{Ours (MASCOT)}            & 0.8772 (0.8667, 0.8881) & 0.5441 (0.9333, 0.3840) \\
\bottomrule
\end{tabular}
\end{small}
\end{center}
\caption{Comparative results on the Visual Genome dataset (\texttt{VG\_hour}) for both diversity increasing and diversity decreasing tasks. The best and second-best overall harmonic mean scores among diversification methods are highlighted in bold and underlined, respectively.}
\label{tab:vg_hour_results}
\end{table*}

\paragraph{I1M\_geo.} Incidents1M is a large-scale geographic benchmark (26K images) where locations are approximated from domain IP geolocation rather than EXIF GPS, introducing coordinate noise that affects all methods equally. Unlike VG and PP where each query has one or very few ground-truth positives, I1M queries (disaster and location types) can match many images in the corpus; we therefore report MAP rather than R@10, as MAP better captures ranking quality across multiple positives per query. MS-DPP leads on both increase (0.7996) and decrease (0.5587) tasks, with \mascot{} competitive on increase (0.7892) but trailing on decrease (0.4973). The I1M geographic attribute is fundamentally different from the PixelProse setting: because locations are derived from server IPs rather than image capture coordinates, many images from the same news domain share identical or near-identical coordinates, creating dense geographic clusters. This cluster structure favors manifold repulsion on increase tasks (pushing apart clustered domains) and makes decrease enforcement difficult for coverage-based methods, as many images naturally fall into the same bins regardless of soft-binning bandwidth. The gap between \mascot{} and MS-DPP on this dataset is therefore partly an artifact of the approximated geographic metadata rather than a fundamental limitation of the coverage formulation.

\begin{table*}[!h]
\begin{center}
\begin{small}
\begin{tabular}{lcc}
\toprule
& \textbf{I1M\_geo (Increase)} & \textbf{I1M\_geo (Decrease)} \\
\textbf{Method} & \textbf{HM$\uparrow$(MAP$\uparrow$, DM$\uparrow$)} & \textbf{HM$\uparrow$(MAP$\uparrow$, DM$\uparrow$)} \\
\midrule
BLIP-2 (Base)             & \underline{0.7955} (0.7442, 0.8544) & 0.4566 (0.7442, 0.3294) \\
\midrule
Clustering                & 0.7685 (0.6900, 0.8672) & 0.4543 (0.6859, 0.3396) \\
MMR                       & 0.7824 (0.7134, 0.8663) & 0.4251 (0.7134, 0.3028) \\
k-DPP                     & 0.7884 (0.7103, 0.8857) & 0.4621 (0.7055, 0.3436) \\
\midrule
MS-DPP                    & \textbf{0.7996} (0.7391, 0.8711) & \textbf{0.5587} (0.6153, 0.5115) \\
MS-DPP+TN                 & 0.7902 (0.7164, 0.8810) & 0.5484 (0.6806, 0.4592) \\
MS-DPP+TN+TVMS            & 0.7884 (0.7234, 0.8663) & \underline{0.5541} (0.6578, 0.4786) \\
\midrule
\textbf{Ours (Prob-Coverage/Both Ablated)} & 0.7696 (0.6788, 0.8884) & 0.5450 (0.6781, 0.4555) \\
\textbf{Ours (w/o Normalization)}          & 0.7919 (0.7178, 0.8830) & 0.4941 (0.6521, 0.3977) \\
\textbf{Ours (Uniform Binning)}            & 0.7952 (0.7333, 0.8685) & 0.5424 (0.6781, 0.4520) \\
\textbf{Ours (MASCOT)}                     & 0.7892 (0.7160, 0.8790) & 0.4973 (0.6949, 0.3871) \\
\bottomrule
\end{tabular}
\end{small}
\end{center}
\caption{Comparative results on the I1M dataset (\texttt{I1M\_geo}) for both diversity increasing and diversity decreasing tasks. The best and second-best overall harmonic mean scores among diversification methods are highlighted in bold and underlined, respectively.}
\label{tab:i1m_geo_results}
\end{table*}

\paragraph{SkyScript\_geo.} SkyScript is a remote-sensing dataset where all images are overhead satellite views with high visual similarity. The near-zero HM scores across all methods (ranging from 0.0355 to 0.0470) reflect the extreme difficulty of the retrieval task rather than a failure of diversification: the base BLIP-2 retriever achieves R@10 of only 0.0242, indicating that semantic matching itself is unreliable in this domain. Under these conditions, diversification methods have almost no relevant signal to redistribute and differences between methods are negligible. \mascot{} attains the numerically highest overall score (0.0458 vs.\ the baseline's 0.0454), but given the near-zero absolute scale, we read this as parity rather than a meaningful advantage: when the base retriever cannot match images to captions, no re-ranker can manufacture relevance.

\begin{table*}[!h]
\begin{center}
\begin{small}
\begin{tabular}{lcc}
\toprule
& \textbf{SkyScript\_geo (Increase)} & \textbf{SkyScript\_geo (Decrease)} \\
\textbf{Method} & \textbf{HM$\uparrow$(R@10$\uparrow$, DM$\uparrow$)} & \textbf{HM$\uparrow$(R@10$\uparrow$, DM$\uparrow$)} \\
\midrule
BLIP-2 (Base)   & \textbf{0.0470} (0.0242, 0.8218) & \underline{0.0454} (0.0242, 0.3649) \\
\midrule
Clustering      & 0.0355 (0.0181, 0.8431) & 0.0381 (0.0200, 0.3811) \\
MMR             & 0.0387 (0.0198, 0.8451) & 0.0373 (0.0198, 0.3176) \\
\midrule
MS-DPP          & 0.0446 (0.0229, 0.8448) & 0.0444 (0.0236, 0.3649) \\
MS-DPP+TN+TVMS  & 0.0383 (0.0196, 0.8574) & 0.0433 (0.0226, 0.5252) \\
\midrule
\textbf{Ours (Prob-Coverage/Both Ablated)} & 0.0388 (0.0199, 0.8325) & 0.0447 (0.0238, 0.3705) \\
\textbf{Ours (MASCOT)}                     & \underline{0.0469} (0.0241, 0.8224) & \textbf{0.0458} (0.0244, 0.3821) \\
\bottomrule
\end{tabular}
\end{small}
\end{center}
\caption{Comparative results on the SkyScript dataset (\texttt{SkyScript\_geo}) for both task directions. The best and second-best overall harmonic mean scores among diversification methods are highlighted in bold and underlined, respectively.}
\label{tab:skyscript_geo_results}
\end{table*}

\section{Implementation Details}
\label{sec:appendix_implementation}
 
\subsection{Composite Attribute Handling}
For composite-attribute tasks such as \texttt{PP\_geo\_hour}, $\mathcal{U}$ is formed by concatenating per-attribute IU sets: $\mathcal{U} = \mathcal{U}_{\text{geo}} \oplus \mathcal{U}_{\text{time}}$, yielding a flat bin space of size $|\mathcal{U}_{\text{geo}}| + |\mathcal{U}_{\text{time}}|$ (e.g., $400 + 24 = 424$ for grid size 20). Each sub-space uses its own Gaussian kernel parameterized by $\sigma_{\text{geo}}$ and $\sigma_{\text{time}}$ respectively, and the soft-assignment matrix $p \in \mathbb{R}^{N \times |\mathcal{U}|}$ is the horizontal concatenation of the per-attribute matrices. The bin importance $\Omega$ and coverage penalty are then computed jointly over this combined space. 

Our main composite tasks apply a single shared direction $d$ across the metadata attributes under refinement. The CDR-CA formulation of \citet{sogi2025msdppsmultisourcedeterminantalpoint} is more general, permitting a distinct direction per attribute; because \mascot{}'s coverage term operates over a partitioned Information Unit space $\mathcal{U} = \mathcal{U}_{\text{geo}} \oplus \mathcal{U}_{\text{time}}$ with no bin-level cross term between attributes, the directional sign can be applied independently within each attribute's coverage sub-term. We implement and evaluate this per-attribute (mixed-direction) setting in Appendix~\ref{sec:appendix_mixed}.

\textbf{MS-DPP+TN+TVMS} is the full Tangent Normalization variant of MS-DPP~\cite{sogi2025msdppsmultisourcedeterminantalpoint}. It applies a two-stage normalization on the SPD manifold:
\begin{enumerate}
    \item \textbf{TN on TVs}: each attribute's tangent vector $\log \mathbf{S}_i$ is normalized by its own Frobenius norm, preventing any single attribute from dominating the weighted sum. (MS-DPP+TN)
    \item \textbf{TN on M}: the unified tangent vector $\log \mathbf{M}'$ is further re-normalized by the Frobenius norm of $\log \mathbf{R}$ (the relevance matrix), aligning the scale of the unified similarity matrix with the relevance signal. (MS-DPP+TN+TVMS)
\end{enumerate}

\subsection{Mixed-Direction Control}
\label{sec:appendix_mixed}
The composite tasks in the main paper apply one direction to all attributes jointly. A practically relevant generalization is \emph{mixed-direction} control. For instance, spreading results across geography while concentrating them in time (geo$\uparrow$+time$\downarrow$), or the converse. Because \mascot{}'s combined space $\mathcal{U} = \mathcal{U}_{\text{geo}} \oplus \mathcal{U}_{\text{time}}$ is partitioned and the directional sign in Algorithm~\ref{alg:mascot} is applied within each attribute's coverage sub-term, per-attribute direction requires only that each sub-space carry its own sign; the change is a keyword argument to the search routine. We verify by bit-identical regression tests that the $[\,\uparrow,\uparrow\,]$ and $[\,\downarrow,\downarrow\,]$ mixed paths reduce exactly to the single-direction implementation, and we extend MS-DPP with an analogous per-attribute sign vector as a baseline.

Table~\ref{tab:mixed} reports both configurations on \texttt{PP\_geo\_hour} ($K=10$). \mascot{} outperforms MS-DPP in both settings: on geo$\uparrow$+time$\downarrow$ it attains R@10 = 0.960 against 0.916, and on geo$\downarrow$+time$\uparrow$, 0.907 against 0.903. The saturation mechanism thus generalizes to asymmetric constraints: because there is no bin-level cross term between attributes, increasing diversity along one attribute and decreasing it along another coexist within a single objective.

\begin{table}[!h]
\centering
\footnotesize
\setlength{\tabcolsep}{4pt}
\begin{tabular}{llccccc}
\toprule
\textbf{Method} & \textbf{Dir.} & $\lambda$ & \textbf{R@10} & \textbf{geo-V} & \textbf{time-V} & \textbf{HM} \\
\midrule
\mascot{} & g$\uparrow$t$\downarrow$ & 0.3 & \textbf{0.960} & 0.907 & 0.791 & 0.934 \\
\mascot{} & g$\downarrow$t$\uparrow$ & 0.8 & \textbf{0.907} & 0.941 & 0.865 & 0.919 \\
\midrule
MS-DPP    & g$\uparrow$t$\downarrow$ & 0.9 & 0.916 & 0.973 & 0.797 & 0.924 \\
MS-DPP    & g$\downarrow$t$\uparrow$ & 0.9 & 0.903 & 0.836 & 0.837 & 0.906 \\
\bottomrule
\end{tabular}
\caption{Mixed-direction results on \texttt{PP\_geo\_hour} (test split, $K=10$).
The \textbf{geo-V} and \textbf{time-V} columns are per-attribute diagnostic
Vendi scores, each normalised as $(V-1)/(K-1)$ on its own feature block
(GPS pair for geo, circular $[\sin,\cos]$ for time). The \textbf{HM} column is
HM(R@10, $\bar{V}$), where $\bar{V}$ is the harmonic-mean Vendi over the
appearance and combined-metadata channels produced by the standard evaluation
pipeline, the same quantity reported as DM in Tables~\ref{tab:decrease}
and~\ref{tab:increase}, but at $K=10$ and without the decrease-direction
$1-x$ transform, since these configurations are run with a top-level
direction of increase. Values are therefore not directly comparable to
Tables~\ref{tab:decrease}--\ref{tab:increase}, and $\bar{V}$ is not the mean of
the two columns shown. \mascot{} outperforms MS-DPP on R@10 and HM in both
configurations.}
\label{tab:mixed}
\end{table}

\subsection{Appearance-Diversity Diagnostic}
\label{sec:appendix_appearance}
A natural objection to \mascot{} is that, unlike MS-DPP, it does not model image appearance as an attribute: MS-DPP includes an appearance similarity term by default, whereas \mascot{} operates only over geo-temporal metadata, although appearance enters the evaluation metric through the DM's appearance channel (Appendix~\ref{sec:appendix_normalization}). If the diversity of appearance mattered independently, \mascot{} should perform poorly. Table~\ref{tab:appearance} tests this by measuring \texttt{img\_vendi}, the Vendi Score over BLIP-2 image embeddings of the selected top-20 set, at the validation-HM-best operating points on \texttt{PP\_geo\_hour}.

\begin{table}[!h]
\centering
\begin{small}
\begin{tabular}{llccc}
\toprule
\textbf{Direction} & \textbf{Method} & \textbf{img\_vendi} & \textbf{DM} & \textbf{R@10} \\
\midrule
Increase & BLIP-2 (base)   & 0.9364 & 0.9142 & 0.9737 \\
Increase & MS-DPP+TN+TVMS  & 0.9400 & 0.9306 & 0.9711 \\
Increase & \mascot{}       & 0.9429 & 0.9435 & 0.8356 \\
\midrule
Decrease & BLIP-2 (base)   & 0.9364 & 0.1656 & 0.9737 \\
Decrease & MS-DPP+TN+TVMS  & 0.9395 & 0.2066 & 0.9021 \\
Decrease & \mascot{}       & 0.9357 & 0.1881 & 0.9410 \\
\bottomrule
\end{tabular}
\end{small}
\caption{Appearance diversity (\texttt{img\_vendi}) alongside metadata DM and R@10 at val-HM-best operating points on \texttt{PP\_geo\_hour}. DM and R@10 reproduce the corresponding entries of Tables~\ref{tab:increase} and~\ref{tab:decrease}.}
\label{tab:appearance}
\end{table}

The objection does not hold empirically. Across both directions and all three methods, \texttt{img\_vendi} clusters within a band of $0.0072$ (0.9357--0.9429), barely distinguishable from the unconstrained BLIP-2 baseline's own 0.9364. On decrease, \mascot{}'s \texttt{img\_vendi} (0.9357) reaches 99.6\% of MS-DPP+TN+TVMS's (0.9395); on increase it is marginally higher. The $N=200$ candidate pool, already filtered for semantic relevance, is itself appearance-diverse, so any selection over a structurally distinct attribute inherits comparable visual variety as a side effect. \mascot{}'s metadata-driven concentration groups selections by geographic cluster and hour, which tends to draw visibly distinct scenes for the same reason that distinct places and times tend to look different, enough to match MS-DPP's explicit appearance term without modeling it.

This also bears on how DM itself should be read. Because DM combines an
appearance and a metadata channel, a method could in principle raise DM by
diversifying appearance rather than metadata. The narrow band above rules this
out on our benchmarks: the appearance channel is effectively constant across
methods, so differences in DM reflect differences in metadata concentration.

\subsection{Generality Beyond Geo-Temporal Metadata}
\label{sec:generality}
The \mascot{} objective is agnostic to what an Information Unit represents, requiring only a soft-binning kernel over the attribute. To test generality beyond geography and time, we induced ten discrete IUs from $k$-means clusters of BLIP-2 image embeddings on PixelProse and used cluster identity as the attribute, with no change to the objective. \mascot{} improved DM from 0.815 to 0.869 at R@10 = 0.98 on increase, and from 0.420 to 0.689 at R@10 = 0.96 on decrease. Because any metadata admitting a soft-binning kernel is supported, the framework extends to learned or semantic attributes without modification.

\subsection{Relevance-Aligned Initialization: Empirical behavior}
\label{sec:appendix_zero_shot}

Section~\ref{sec:penalty} notes that the coverage gain is non-zero at $m=1$. Instrumenting the greedy loop confirms this directly: on \texttt{PP\_geo\_hour} decrease at the operating point of Table~\ref{tab:decrease}, $\Delta_{\text{cov}}$ at the first step ranges over $[1.09, 12.21]$ with mean $7.46$, and is non-zero for all 797 queries. No guarantee of top-1 preservation exists, and none is observed.

Because $\Omega(u, q) = \max_{j} p(u, j)\,\hat{R}(j, q)$ is constructed from normalized relevance, bins containing high-relevance images receive high $\Omega$, and the most relevant image therefore also carries a high coverage gain. The consequence depends on direction. For $d = +1$ the relevance and coverage terms favour the same candidate and the top-ranked result is largely retained. For $d = -1$ the coverage term enters as a penalty, so the same correlation makes the most relevant image the most heavily penalized; top-1 displacement is the expected behavior of the objective rather than a failure of it. Consistent with this, \mascot{}'s first selection matches the base retriever's on 84.4\%, 76.0\% and 65.8\% of queries on the three decrease tasks, with agreement falling as $\lambda$ rises, and its R@1 falls below the unconstrained baseline (0.7905) on all six PixelProse task-directions --- by more on decrease (0.7202, 0.6688, 0.6048) than on increase (0.7578, 0.7227, 0.7553).

We further characterize the displacements. For queries where the base retriever's top-1 is the ground-truth image and \mascot{}'s differs, we classify each case by whether the two images occupy different primary Information Units. Across the three decrease tasks, 353 of 356 displacements (99.2\%) move to a different primary bin, consistent with concentration into the target metadata region rather than indiscriminate loss of relevance. This is a consistency check rather than a validation: because concentration into a target bin is what the objective rewards when $d = -1$, a high rate is close to expected.

\subsection{Diversity Metric and Overall Score Normalization}
\label{sec:appendix_normalization}

We adopt the diversity metric of \citet{sogi2025msdppsmultisourcedeterminantalpoint}
without modification. Because its two-channel structure matters for interpreting every DM value we report, we restate it here. For a selected subset of size $K$, the Vendi Score $V$ is computed separately on two feature sets: the base VLM's image embeddings (the \emph{appearance} channel) and the task's metadata features (the \emph{metadata} channel). Each is normalized to $[0,1]$ as
\begin{equation}
    \tilde{V} = \frac{V - 1}{K - 1}.
\end{equation}
For decrease tasks the transform $1 - \tilde{V}$ is applied to the
\emph{metadata channel only}; the appearance channel is never inverted. The
two per-query values are then combined by a harmonic mean, and the Diversity
Metric is the harmonic mean of that quantity across queries:
\begin{equation}
    \text{DM} = \mathrm{HM}_{q}\!\left(
      \mathrm{HM}\big(\tilde{V}^{\text{app}}_{q},\;
                      \tilde{V}^{\text{meta}\prime}_{q}\big)\right).
\end{equation}
Because the harmonic mean is dominated by its smaller argument, and because
the appearance channel is near-constant across methods
(Appendix~\ref{sec:appendix_appearance}), DM on decrease tasks varies almost
entirely with the metadata channel; the appearance channel acts as a
compressive factor rather than a discriminative one. The Overall Score is the
harmonic mean of R@10 (or MAP for I1M) and DM. Baseline Vendi scores differ
across datasets and attribute types, so absolute HM values should be compared
within a task across methods, not across tasks.

\subsection{Runtime Analysis}
\label{sec:appendix_runtime}

Our submission reported a per-query latency of 1266\,ms for \mascot{}, dominated by a sequential Python greedy loop rather than the underlying computation. A vectorized reimplementation, \texttt{search\_vec()}, runs the identical objective end-to-end in 2.99\,ms at $N=200$, $K=20$, $|\mathcal{U}|=424$ on an NVIDIA RTX A6000, with additional GPU memory below 2\,MB per query. \texttt{search\_vec()} is numerically identical to the sequential path in float32. Accuracy results throughout this paper use the sequential implementation, while runtime is measured on \texttt{search\_vec()}.

\begin{table}[!h]
\centering
\begin{small}
\begin{tabular}{lc}
\toprule
\textbf{Method} & \textbf{Time (ms/query)} \\
\midrule
MS-DPP                          & 26.15 \\
\mascot{} (sequential)          & 1266.21 \\
\mascot{} (vectorized)          & \textbf{2.99} \\
\bottomrule
\end{tabular}
\end{small}
\caption{Per-query wall-clock time at $N=200$, $K=20$ on an NVIDIA RTX A6000. MS-DPP uses a single batched \texttt{scipy.linalg.eigh} (LAPACK) call. The vectorized \mascot{} implementation is numerically identical to the sequential one and reduces latency by roughly $420\times$.}
\label{tab:runtime}
\end{table}

The theoretical complexity of the greedy loop is $O(K \cdot N \cdot |\mathcal{U}|)$, and the established object-level and iteration-level pruning for probabilistic coverage \cite{xu2014efficient} applies directly. Both \mascot{} and MS-DPP are post-hoc re-rankers applied once per query.

\section{Code and Reproducibility}
\label{sec:appendix_code}

The complete implementation of \mascot{}, including all baselines, evaluation scripts, sensitivity analyses, and post-submission extensions, is available at:
\begin{center}
\url{https://github.com/AaryanSharma/MASCOT}
\end{center}
The repository includes the greedy subset selection implementation, preprocessing for all four benchmarks, the vectorized runtime benchmark, and scripts to reproduce every table and figure.

\begin{figure*}[t]
    \centering
    \includegraphics[width=\linewidth]{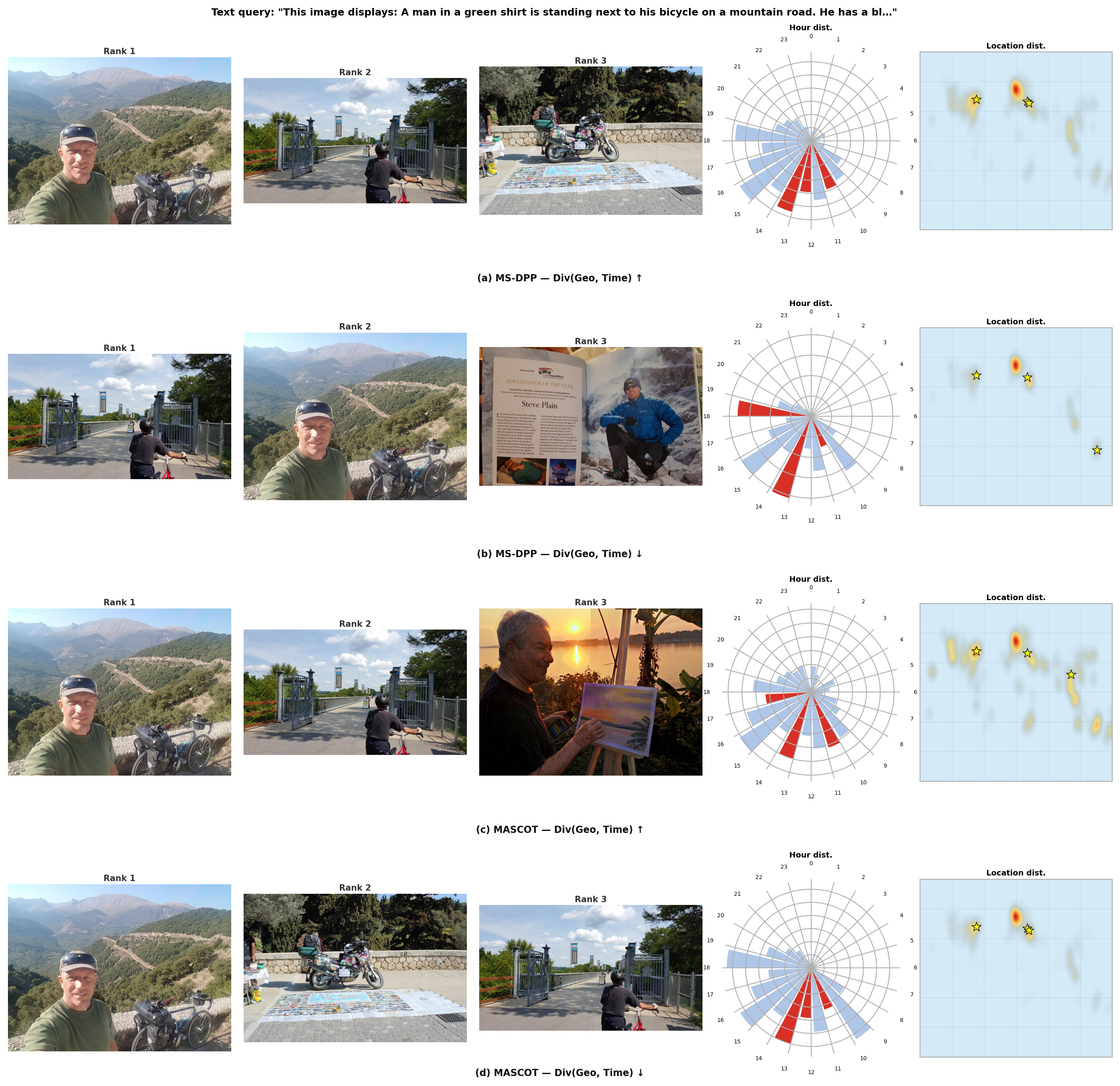}
    \caption{
    Qualitative comparison of retrieval under composite-attribute refinement on \texttt{PP\_geo\_hour}. Each row displays the top-3 retrieved images along with attribute distributions (geographic heatmaps and temporal polar histograms).
    \textbf{Metadata Visualizations:} In the polar histograms, the \textbf{red bars} indicate the exact hour the top-3 images were taken.
    In the geo-heatmaps, the \textbf{yellow stars} indicate their exact GPS locations.
    }
    \label{fig:qualitative_results1}
\end{figure*}

\begin{figure*}[t]
    \centering
    \includegraphics[width=\linewidth]{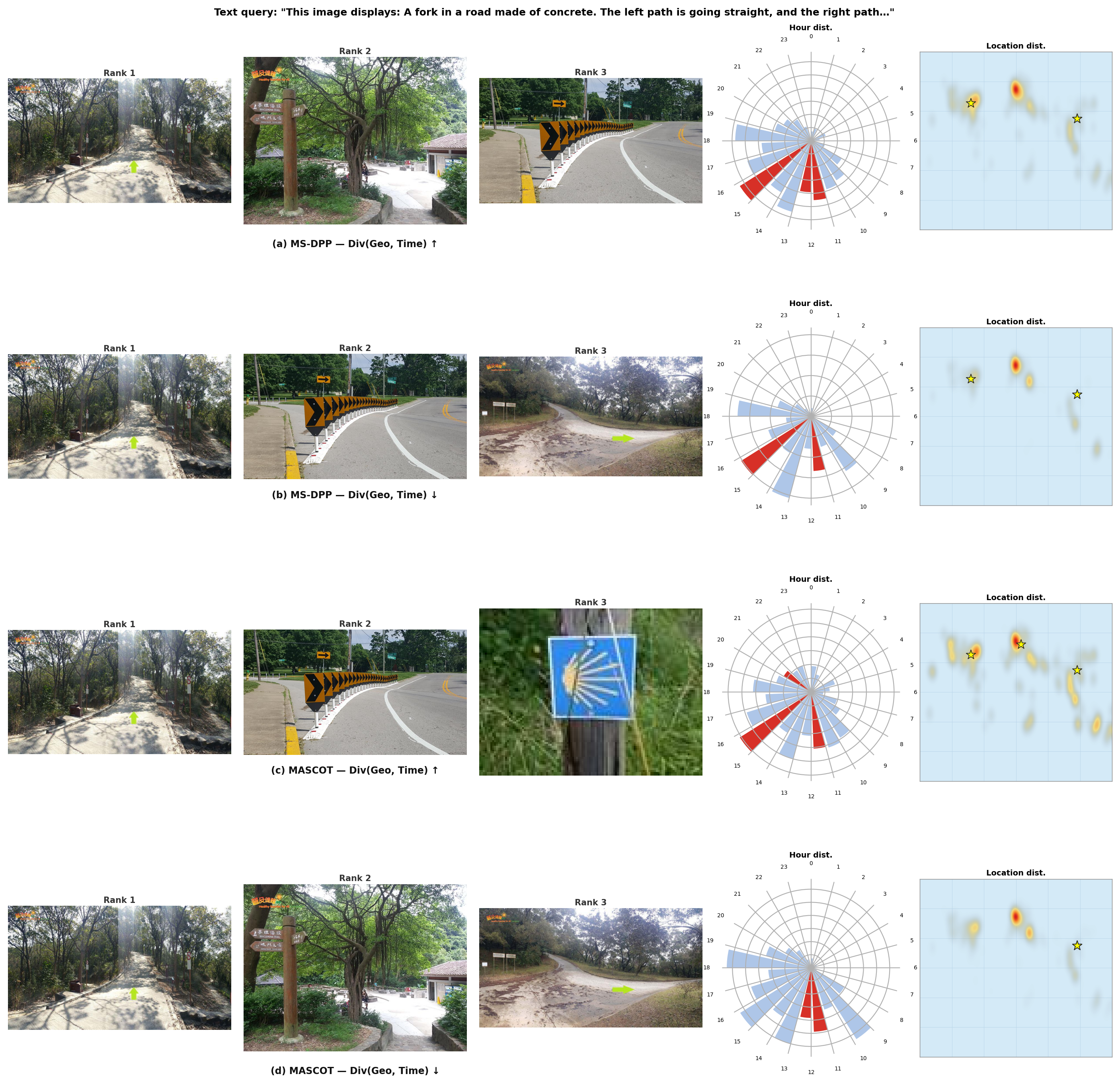}
    \caption{
    Qualitative comparison of retrieval under composite-attribute refinement on \texttt{PP\_geo\_hour}. Each row displays the top-3 retrieved images along with attribute distributions (geographic heatmaps and temporal polar histograms).
    \textbf{Metadata Visualizations:} In the polar histograms, the \textbf{red bars} indicate the exact hour the top-3 images were taken.
    In the geo-heatmaps, the \textbf{yellow stars} indicate their exact GPS locations.
    }
    \label{fig:qualitative_results2}
\end{figure*}

\end{document}